\documentclass[preprint,10pt]{aastex}

\newcommand\sqd{{deg$^{2}$}}
\newcommand\etal{{\it et al. }}
\newcommand\kms{km~s$^{-1}$~}
\newcommand\msun{$M_\odot$}

\def\ncat{730}
\def\ndet{586}
\def\nprior{120}
\def\nhvc{24}
\def\nooc{41}
\def\be{\begin{equation}}
\def\ee{\end{equation}}

\shorttitle{ALFALFA Catalog of Northern Virgo Region}
\shortauthors{Giovanelli et al.}

\begin{document}
%\hskip 3.5in{Version 2.1 \hskip 10pt 04Dec2006}
\title{The Arecibo Legacy Fast ALFA Survey: \\
       III. HI Source Catalog of the Northern Virgo Cluster Region  
}

\author {Riccardo Giovanelli\altaffilmark{1,2}, Martha P. Haynes\altaffilmark{1,2}, 
Brian R. Kent\altaffilmark{1}, Am\'elie Saintonge\altaffilmark{1},
Sabrina Stierwalt\altaffilmark{1},
Adeel Altaf\altaffilmark{3},
Thomas Balonek\altaffilmark{4,2},
Noah Brosch\altaffilmark{5},  
Shea Brown\altaffilmark{6},
Barbara Catinella\altaffilmark{7}, 
Amy Furniss\altaffilmark{8},
Josh Goldstein\altaffilmark{3}, 
G. Lyle Hoffman\altaffilmark{3},
Rebecca A. Koopmann\altaffilmark{9,2}, 
David A. Kornreich\altaffilmark{8}
Bilal Mahmood\altaffilmark{9}
Ann M. Martin \altaffilmark{1},
Karen L. Masters \altaffilmark{10},
Arik Mitschang\altaffilmark{8},
Emmanuel Momjian\altaffilmark{7}, 
Prasanth H. Nair\altaffilmark{11},
Jessica L. Rosenberg\altaffilmark{12},
Brian Walsh\altaffilmark{4}
}
%\affil{Department of Astronomy and National Astronomy and Ionosphere Center
%\altaffilmark{1},\\
%Space Sciences Bldg., Cornell University, Ithaca, NY 14853\\ 
%{\it e--mail:} riccardo@astro.cornell.edu,  haynes@astro.cornell.edu}

\altaffiltext{1}{Center for Radiophysics and Space Research, Space Sciences Building,
Cornell University, Ithaca, NY 14853. {\it e--mail:} riccardo@astro.cornell.edu,
haynes@astro.cornell.edu, bkent@astro.cornell.edu, amartin@astro.cornell.edu, 
amelie@astro.cornell.edu, sabrina@astro.cornell.edu}

\altaffiltext{2}{National Astronomy and Ionosphere Center, Cornell University,
Space Sciences Building,
Ithaca, NY 14853. The National Astronomy and Ionosphere Center is operated
by Cornell University under a cooperative agreement with the National Science
Foundation.}

\altaffiltext{3}{Hugel Science Center, Lafayette College, Easton, PA 18042.
{\it e--mail:} hoffmang@lafayette.edu, goldstj@lafayette.edu, altafa@lafayette.edu}

\altaffiltext{4}{Dept. of Physics \& Astronomy, Colgate University, Hamilton, NY 13346.
{\it e--mail:} tbalonek@mail.colgate.edu, bwalsh@bu.edu}

\altaffiltext{5}{The Wise Observatory \& The School of Physics and Astronomy, 
Raymond \& Beverly Sackler Faculty of Exact Sciences, Tel Aviv University, Israel.
{\it e--mail:} noah@wise.tau.ac.il}

\altaffiltext{6}{Astronomy Dept., U. of Minnesota, 116 Church St. SE, Minneapolis, 
MN 55455. {\it e--mail:} brown@astro.umn.edu}

\altaffiltext{7}{Arecibo Observatory, National Astronomy and Ionosphere Center,
Arecibo, PR 00612. {\it e--mail:} bcatinel@naic.edu, emomjian@naic.edu}

\altaffiltext{8}{Dept. of Physics, Humboldt State University, Arcata, CA  95521.
{\it e--mail:} dak24@humboldt.edu}

\altaffiltext{9}{Dept. of Physics \& Astronomy, Union College, Schenectady, NY 12308.
{\it e--mail:} koopmanr@union.edu, mahmoodb@union.edu}

\altaffiltext{10}{Harvard--Smithsonian Center for Astrophysics, 60 Garden St. MS 65, 
Cambridge MA 02138--1516. {\it e--mail:} kmasters@cfa.harvard.edu}

\altaffiltext{11}{Astronomy Dept., Swain West 319, Indiana University, Bloomington, IN  47405. 
phnair@astro.indiana.edu}

\altaffiltext{12}{Harvard--Smithsonian Center for Astrophysics, 60 Garden St. MS 65, 
Cambridge MA 02138--1516, currently at George Mason University. {\it e--mail:} jrosenb4@gmu.edu}

%\hsize 6.5 truein
\begin{abstract}
We present the first installment of HI sources extracted 
from the Arecibo Legacy Fast ALFA (ALFALFA) extragalactic survey, initiated
in 2005. Sources have been extracted from  
3-D spectral data cubes exploiting a matched filtering technique and 
then examined interactively to yield
global HI parameters. A total of \ncat~ HI detections are catalogued within the 
solid angle $11^h44^m <$ R.A.(J2000) $< 14^h00^m$ and 
$+12^\circ <$ Dec.(J2000) $<+16^\circ$, and redshift range $-1600$ \kms 
$~< cz < 18000$ \kms. In comparison, the HI Parkes All-Sky Survey (HIPASS)
detected 40 HI signals in the same region. Optical counterparts are assigned via examination
of digital optical imaging databases. ALFALFA HI detections are reported for three 
distinct classes of signals: (a) detections, typically with S/N $>$ 6.5; (b) high velocity clouds
in the Milky Way or its periphery; and (c) signals of lower S/N (to $\sim$ 4.5)
which coincide spatially with an optical object of known similar redshift.
Although this region of the sky has been heavily surveyed by previous targeted 
observations based on optical flux-- or size-- limited 
samples, 69\% of the extracted sources are newly reported HI detections.
The resultant positional accuracy of HI sources is dependent on S/N: it 
averages 24\arcsec ~(20\arcsec ~median) for all sources with S/N $>$ 6.5 and is of order 
$\sim$17\arcsec ~(14\arcsec ~median) for signals with S/N $>$ 12. The median
redshift of the sample is $\sim$7000 \kms ~and its
distribution reflects the known local large scale structure including
the Virgo cluster and the void behind it, the A1367-Coma supercluster at 
$cz \sim$7000 \kms~ and a third more distant overdensity at $cz \sim$13000 \kms.
Distance uncertainties in and around the Virgo cluster perturb the derived 
HI mass distribution. Specifically, an apparent deficiency
of the lowest HI mass objects can be attributed, at least in part, to the incorrect
assignment of some foreground objects to the cluster distance. Several
extended HI features are found in the vicinity of the Virgo cluster. A
small percentage (6\%) of HI detections have no identifiable optical counterpart,
more than half of which are high velocity clouds in the Milky Way vicinity; the remaining 17
objects do not appear connected to or associated with any known galaxy. Based on 
these initial results, ALFALFA is expected to fulfill, and even exceed, its
predicted performance objectives in terms of the number and quality of 
HI detections.
\end{abstract}


\section {Introduction}\label{intro}
Because the evolution of the HI content with time promises to be
a powerful tracer of galaxy evolution even into the so-called
``Dark Ages'', it remains imperative to understand fully the
true census of HI-bearing objects at $z \sim 0$. 
As spectroscopic tracers, HI emission line profiles yield not only the
redshift, but also measures of the total HI mass and the radial component
of the rotational velocity via the HI line width, thus providing
quantitative insight into the gas content and total mass. 
At the same time that wide area optical (e.g. Sloan Digital Sky Survey: SDSS;
York \etal ~2000)
and near-infrared (e.g. Two-Micron All-Sky Survey: 2MASS; Skrutskie 
\etal ~2006) surveys have been
cataloging millions of galaxies,
the HI Parkes All-Sky Survey (HIPASS; Barnes \etal ~2001) produced the first
wide area blind extragalactic HI line survey. 
HIPASS covered $\sim$30000 \sqd, producing a final catalog
(Meyer \etal ~2004) of 4315 HI detections south of Dec. = $+2^\circ$,
and another 1002 (Wong \etal ~2006) between $+2^\circ <$ Dec $< +20^\circ$.
Because HIPASS was limited in depth and resolution, it did not sample adequate
volume to yield a ``cosmologically fair'' sample of the universe; the median
redshift of HIPASS sources is $\sim$2800 \kms, near the outer boundary
of the Local Supercluster. Of critical importance for the determination
of the faint-end slope of the HI mass function (HIMF),
only a handful of galaxies of very low HI mass were detected, and 
those were so near the Milky Way that distance uncertainties make their
HI masses likewise highly uncertain (Masters \etal ~2004).  

HI line surveys are of special importance when combined with
optical/IR surveys so that both the stellar and gaseous components are
sampled. As was shown also by earlier studies such as the Arecibo Dual Beam 
Survey (ADBS: Rosenberg \& Schneider 2002), blind HI surveys detect a
population of low surface brightness, gas rich objects which are
often missed by magnitude limited optical and near-IR samples. 
Likewise, not all galaxies exhibit detectable HI line emission, and
HI surveys are limited to some degree by their positional
accuracy and angular resolution since the telescopes involved 
have relatively large beam sizes. 
A true accounting of the extragalactic census must incorporate
full sampling of the population of gas-rich objects sampled by
HI surveys in complement to the stellar-rich ones detected
by surveys like SDSS and 2MASS.

Based both on the
success of HIPASS and its limitations, we have initiated a
``second generation'' HI blind survey, the Arecibo Legacy
Fast ALFA (ALFALFA) survey, which exploits the availability of the 
new multi-beam Arecibo L-band Feed Array (ALFA) on the
305~m antenna. ALFALFA will require $\sim$ 4000 hours
of telescope time to survey $\sim$7000 \sqd ~of the high galactic 
latitude sky visible from Arecibo (Giovanelli
\etal ~2005a: Paper I). HIPASS used a 13
beam receiver on the Parkes 62-m telescope (15.5\arcmin ~HPBW); ALFALFA 
exploits a similar 7-feed receiver array on the 305-m Arecibo telescope
($\sim$3.5\arcmin). The survey design of ALFALFA exploits 
the superior collecting area and angular resolution of the Arecibo
telescope as well as the broader bandwidth and spectral resolution
of its digital correlator. Furthermore, ALFALFA makes use of automated signal
detection techniques to produce source catalogs unbiased by the width
of the HI signal (Saintonge 2007). Designed to surpass significantly the HI survey benchmark 
established by HIPASS, ALFALFA improves on its predecessor by a factor 
of $\sim 8$ in sensitivity,
$\sim 4$ in angular resolution, $\sim 3$ in spectral resolution and
$\sim 1.6$ in spectral coverage. The added depth of ALFALFA
will allow it to sample a ``fair'' volume of the Universe. In particular, ALFALFA will sample well
the volume corresponding to the so-called ``convergence depth'' 
(Giovanelli \etal ~1998; Dale \& Giovanelli 2000) which contributes most
of the peculiar velocity of the Milky Way with respect to the Cosmic
Microwave Background.

Simulations presented in Paper I predicted that
ALFALFA will detect some 20,000 extragalactic HI line sources, from
very nearby low mass dwarfs to massive spirals at $z \sim 0.06$.
The survey is designed specifically to determine robustly 
the HIMF in the local universe and will at the 
same time provide a census of HI in the surveyed sky area, making 
it especially useful in synergy with other wide area surveys such 
as SDSS, 2MASS, GALEX, ASTRO--F, etc. The full scope and goals of the
survey are described further in Paper I.

HI line surveys yield three principal observational parameters: the
integrated HI line flux, the systemic redshift and the Doppler
line width. For resolved objects, they can also yield an estimate
of the HI size, and if the source fills the beam, the HI column
density. ALFALFA is expected to resolve about 500 nearby galaxies;
other objects with exceptionally extended HI distributions will
also be mapped. Its positional accuracy (see Section \ref{pos}) allows
identification of the optical counterparts of the HI detections
by immediate cross reference with the large optical/IR imaging databases.
The most interesting objects will perhaps be the isolated HI sources which
may {\it not} have optical counterparts, the so-called ``dark
galaxies''.

The ALFALFA survey was initiated in February 2005. Since then, we
have conducted observations regularly and anticipate allocations
of 700--900 hours of telescope time per year. In order to provide
timely information to the community, a public web site 
{\it http://egg.astro.cornell.edu/alfalfa} is updated regularly
to provide survey plans and status information. Because of the
2-pass drift mode observing strategy (Paper I), complete spectral 
coverage combines many separate datasets obtained over 
observing periods spread over many months. Final 3-D cubes from which
signals can be reliably extracted cannot be produced until the entire 
dataset covering a targeted region is acquired. In fact, the 
first-year datasets were rather incomplete, so that construction of 
full-coverage 3-D data cubes has been possible only recently.

Given the area of sky visible to the Arecibo telescope,
a prime target for ALFALFA is the Virgo cluster and its 
surroundings. Although the spirals 
in the core of the Virgo cluster are strongly HI deficient
(Davies \& Lewis 1973; Chamaraux \etal ~1980; Giovanelli \& Haynes 1983; Solanes
\etal ~2002), the cluster periphery contains many known optical
late-type galaxies. ALFALFA is specifically designed to detect
objects of relatively low HI mass, $\sim 3 \times 10^7$\msun,
at the Virgo distance. Hence, the Virgo region of the sky was
targeted by ALFALFA from the beginning; its central region has been
the first zone to be surveyed completely. 
In this paper, we present a catalog of HI sources extracted
from two constant declination strips of 3-D grids, covering 
a swath $4^\circ$ wide in Declination $+12^\circ <$ Dec.(J2000) 
$<+16^\circ$ and stretching $33^\circ$ in Right 
Ascension between $11^h44^m <$ R.A.(J2000) $< 14^h00^m$.
This area includes the northern section of the Virgo cluster. 
With a complete dataset now in hand for a first installment covering
132 \sqd, this paper presents the first
ALFALFA HI detection catalog, in the spirit of prompt access which 
the legacy nature of the survey promises. A catalog containing 
sources of the southern part of the Virgo cluster region is in preparation
(Kent \etal ~2007).

In Section \ref{obs} we discuss the ALFALFA survey observations and 
data analysis. In Section \ref{cat}, we briefly discuss issues 
related to the data quality of the region sampled by the catalog, 
as well as signal extraction criteria. The contents of the catalog 
are also described. Section \ref{stats} gives an overview of the
statistical properties of the cataloged sample. The
positional accuracy of the HI positions is discussed in Section \ref{pos}.
Section \ref{disc} summarizes the HI detections presented here and what
the results of this first installment predict for the future results
of the full ALFALFA survey. A Hubble constant of 70 \kms~ Mpc$^{-1}$
is used for distance dependent calculations, unless otherwise specified.

\section{Observations and Data Reduction}\label{obs}

ALFALFA uses the 7--feed ALFA receiver system and a spectral line backend
capable of instantaneously producing spectra from the two linear polarizations of
each beam and covering a bandwidth of 100 MHz. The angular
resolution of the survey is given by the elliptical shape of each of ALFA's beams,
$3.3$\arcmin $\times 3.8$\arcmin (Paper I), and the spectral resolution is 25 kHz, which
translates to about 5.5 \kms ~at $cz \sim 0$ \kms ~before spectral smoothing is applied.
ALFALFA surveys the sky with the telescope in ``almost fixed azimuth'' drift mode: the
telescope azimuth arm is placed on the meridian, and the sky drifts by. Small adjustments
of the zenith angle are applied throughout the observing period to maintain the
beam tracks at constant J2000 Dec. Each region of sky
is visited twice, at two epochs spaced by a few months in the Earth's orbit 
about the Sun. No Doppler tracking of the Local Oscillator frequency is employed, 
so that cosmic signals shift in frequency between the two epochs by the difference
in their heliocentric velocities, projected along the line of sight. The resulting coverage
yields parallel tracks of constant J2000 Dec., separated from one another
by 1.05\arcmin. The sampling rate in R.A. is 1 Hz. In 
both coordinates, the sampling is significantly better than the
Nyquist rate. Data-taking for ALFALFA was initiated in 
February 2005 and, in the practical context of time allocation 
at a widely used, multidisciplinary national facility like Arecibo, 
completion of the full survey is projected to require about 6 years.
More technical detail regarding the equipment, observing mode and
sensitivity issues can be found in Paper I.

The ALFALFA data processing scheme has been briefly outlined in Paper I and will
be described in detail elsewhere (Giovanelli \etal ~2007, in preparation; 
Saintonge 2007, submitted).
Full processing of all survey data to level I --- which includes bandpass calibration,
radio frequency interference (hereafter ``rfi'') flagging, continuum
source identification and extraction of drift scans --- is carried out
shortly after data taking, as anticipated in Paper I. However, the production
of 3-D data cubes which fully sample a region of sky requires the completion of
both passes and thus cannot be completed until all data covering that region are 
in hand. In Paper I, we discussed the parceling of the sky in data processing 
units we referred to as ``tiles'',
of $10^m$ in R.A. by $4^\circ$ in Dec. In 
practice, as the processing needs demanded a distributed strategy, 
data units of smaller size than the above mentioned tiles were found to
be compatible with a ``minimum denominator'' computer performance.
The data units thus adopted each cover a sky area of $2.4^\circ \times 2.4^\circ$
in R.A. and Dec.; in order to avoid confusion with the previous definition
of tiles, we refer to regularly gridded data cubes of $2.4^\circ \times 2.4^\circ$ 
and preset centers as ``grids''. Centers of grids that are
adjacent in R.A. are separated from each other by $8^m$; the separation
in Dec. is $2^\circ$. A spatial overlap of about 20\% in each coordinate
allows effective spatial coverage of all sources found in the overlap
regions. In order to maintain the gridded data file size at a limit
comfortably manipulated by modest computer hardware, we break each
spatial grid into four overlapping spectral subgrids of 1024 channels
each, after discarding the bandends. Because of rfi, the spectral resolution
is not reduced at this stage, and, the full information on spectral weights,
as described in Sectin \ref{rfi} below,
is retained. Each 1024 channel 3-D spectral cube has a final size of
380 MB.

\subsection{Access to Data Products}

In the interest of timely access to our data products
by the community, we plan to release source catalogs at the earliest possible
time, consistent with the legacy character of ALFALFA.
The catalog presented here is the first part of an archival database being
developed as a collaborative project with the National Astronomy and Ionosphere
Center and the Cornell Theory Center. The data catalog products will be available 
at {\it http://arecibo.tc.cornell.edu/hiarchive} as a node of the U.S. National
Virtual Observatory\footnote{This research has made use of data obtained 
from or software provided by
the US National Virtual Observatory, which is sponsored by the National
Science Foundation.} (NVO). Included already at that website are catalogs and
spectral data products of targeted single-beam HI observations 
of $\sim$9000 galaxies observed by our group (Springob \etal ~2006)
and the ALFALFA precursor observations (Paper II).
At the time of publication of this paper, the catalog presented here
and access to the spectral profiles associated with the HI detections
will be incorporated into our existing HI digital archive. 
An ongoing development effort aims its focus on the protocol
for long--term public delivery of the 3--D ALFALFA data set through web--based 
access tools. At this time, delivery of the 3-D data is made possible 
through the observing team itself, by direct contact to R.G. or M.P.H..
A major challenge is data volume: each of the 3-D cubes covering the 34 individual 
``grids'' constituting the current catalog require 1.5 GB. 
Allowing access and manipulation of the gridded data publicly will require
the development of distributed computational tools.

In this work, we present a catalog of HI sources extracted
from the ALFALFA grids covering a region stretching from 
$+12^\circ$ to $16^\circ$ in Dec. and from 
$11^h44^m$ to $14^h00^m$ in R.A. For reference to our database,
the denominations of the grids constituting this catalog are 1148+13 to 
1356+13 and 1148+15 to 1356+15, in steps of 8$^m$ in R.A.
The solid angle subtended by this region is $\sim$132 \sqd, 
which is $\sim1.9$\% of the sky to be ultimately surveyed by
ALFALFA. The coverage of the region is complete by the target goals
of ALFALFA, i.e. the region has been sampled by two separate passes
with the ALFA array in drift mode. The average quality of the data is
fair to excellent; details on the data quality at a particular location
can be garnered by consultation with any of the Cornell authors or by
consultation of the ALFALFA website mentioned in Section 1.
This type of information will eventually be accessible
through our NVO--compliant node.

\subsection{Impact of M87}\label{m87}
The surveyed region includes the northern part of the Virgo cluster
and M87, at $RA=+12^h30^m49.4^s$, $Dec=12^\circ23\arcmin28\arcsec$. The 
very strong radio source associated with M87 has a 1.4 GHz continuum flux of 220 Jy, 
which increases the system temperature of the Arecibo telescope at L band by a 
factor of 70 when M87 crosses within the main beam of the telescope. Even when 
M87 is detected only through the near sidelobe structure of the beam, the sensitivity 
decreases drastically both because of the increased system temperature and 
because of the onset of strong spectral standing waves. Effectively, a region of
about $1^\circ$ to 1.5$^\circ$ radius centered on M87 remains inaccessible 
to HI spectroscopy except for the very brightest HI sources, as
we illustrate in Section \ref{stats}.

%FIGURE 1
\begin{figure}[h]
%\figurenum{1}
\plotone{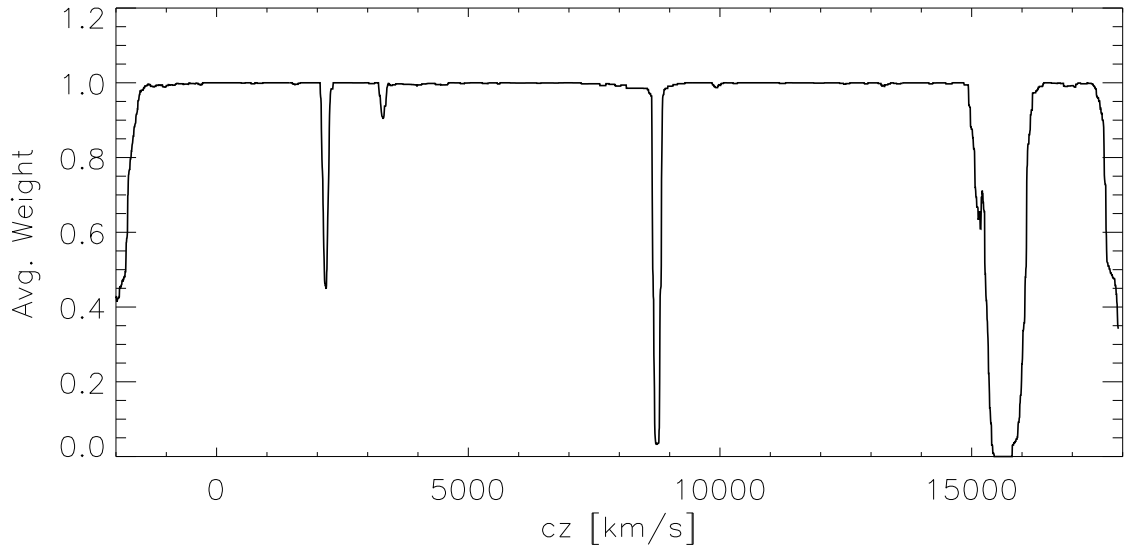}
\caption{Average spectral weights for the region of this catalog,
plotted vs. heliocentric recessional velocity. The steep dropoff at each
end of the spectral domain is an instrumental effect associated with the 
spectrometer bandpass edge. The other strong troughs are associated with 
rfi originating with the San Juan FAA radar operating at 1350 MHz.}
\label{weights.ps}
\end{figure}

\subsection{Radio Frequency Interference}\label{rfi}
In practice, rfi
contaminates certain regions of the observed spectral window.
The data processing pipeline tracks the flagging from visual inspection 
of spectral samples which are
deemed to be contaminated and assigns a weight to each
pixel in 3-D space according to the degree of this flagging.
The quality of the spectral coverage is exemplified in
Figure \ref{weights.ps}, which shows the average ``spectral weight''
over the full region sampled by the present source catalog. A spectral weight of
1.0 indicates good data quality and full utilization of all the
data at that frequency or cosmological recession velocity. We compute
an array of spectral weights, which is a quality estimator, for each 
spatial pixel of the survey. The drops in the spectral weights at both
ends of the graph in Figure \ref{weights.ps} arise from the diminished
sensitivity found at the bandpass edges resulting from instrumental effects.
The deep trough between 15000 and 16000 \kms ~is due to rfi originating
from the Federal Aviation Administration radar operating at 1350 MHz
near the Luiz Mu\~{n}oz Marin airport in San Juan.
The narrower features at 8800 \kms, 3300 \kms ~and 2200 \kms ~are
modulations of the 1350 MHz radar signal occurring within the
spectrometer. While the feature between 15000 and 16000 \kms ~renders 
inaccessible a larger cosmic volume, the feature at 8800 \kms ~affects
a larger number of possible detections, given its spectral location in
vicinity of the peak of the redshift distribution for our catalog.
On average, about 85\% of the spectral bandpass is
clear for astronomical use.

The presence of rfi causes distinct volumes of the universe, which correspond
to the frequency ranges of redshifted HI, to be ``obscured'' to ALFALFA. 
In order to maintain a proper census of the survey volume, we use the 3-D 
``pixel mask'' which records the spectral weights described above.

\section{Source Catalog}\label{cat}

\subsection{Tabulated Data} \label{tab}

Shortly after each observing session, the individual 600-sec drift scans are 
processed to Level I, that is, they are bandpass subtracted and calibrated,
and flagging of channels affected by rfi is performed (see Paper I). 
Once a region of the sky has been fully sampled by ALFALFA, a 3-D grid is then
generated from all of the individual drift scans that cross it. In practice,
the band edges are dropped, and four spectral subgrids, each comprising 1024 
channels, with $\sim 100$ channel overlap of each, are constructed for each  
$2.4^\circ \times 2.4^\circ$
spatial grid. Each spectral subgrid is then
``flat fielded'' (i.e. baselined) and made
ready for source extraction. An automatic signal extractor described in
Saintonge (2007) that operates
in the Fourier domain produces a preliminary catalog of sources to a
specified threshold in signal-to--noise ratio (hereafter S/N). Each 
candidate source is then examined visually and a decision on its inclusion 
in a final catalog, as well as a detailed, interactively obtained set of 
measurements yielding the principal source parameters, is made. A spatially 
integrated spectrum of the source and a postage--stamp
3D section of the grid centered on the source are archived. 

After HI source candidates are inspected, measured and cataloged, 
identification of possible optical counterparts is carried out through
inspection of the Sloan Digital Sky Survey (hereafter SDSS)
\footnote {Funding for the Sloan Digital Sky Survey (SDSS) has been 
provided by the Alfred P. Sloan Foundation, the Participating Institutions, 
the National Aeronautics and Space Administration, the National Science 
Foundation, the U.S. Department of Energy, the Japanese Monbukagakusho, 
and the Max Planck Society. The SDSS Web site is http://www.sdss.org/.
The SDSS is managed by the Astrophysical Research Consortium (ARC) for 
the Participating Institutions. The Participating Institutions are The 
University of Chicago, Fermilab, the Institute for Advanced Study, the 
Japan Participation Group, The Johns Hopkins University, Los Alamos National 
Laboratory, the Max-Planck-Institute for Astronomy (MPIA), the Max-Planck-Institute 
for Astrophysics (MPA), New Mexico State University, University of Pittsburgh, 
Princeton University, the United States Naval Observatory, and the University 
of Washington.}, DSS2 via {\it Skyview}\footnote{{\it Skyview} was developed and 
maintained under NASA ADP Grant NAS5--32068 under the auspices of the High 
Energy Astrophysics Science Archive Research Center at the Goddard Space 
Flight Center Laboratory of NASA.}, NED\footnote{The NASA/IPAC Extragalactic
Database (NED) is operated by the Jet Propulsion Laboratory, California
Institute of Technology, under contract with the National Aeronautics and 
Space Administration.} and our privately maintained data base of extragalactic
sources (the ``AGC'', for ``Arecibo General Catalog''). For the vast majority
of HI source candidates of fair to excellent S/N, a relatively unambiguous
identification of an optical counterpart is possible, thanks to the 
positional accuracy of the HI sources, as described in Section \ref{pos}.
In our catalog we distinguish between 3 kinds of sources: (a) reliable,
extragalactic 
HI candidate sources, down to a S/N threshold of approximately S/N=6.5
(the limit is soft, as described further below); (b) high S/N features of 
low velocity likely to be galactic or peri--galactic High Velocity Clouds (HVCs);
(c) candidate HI sources of lower S/N (approximately 4.5 to 6.5), corroborated 
by the vicinity of a possible optical counterpart of the same redshift. The 
signal extraction process also yields candidate
sources of lower S/N than those listed in the present catalog. A list
of those objects is available upon request. As discussed in Paper I,
the most efficient strategy for confirming follow-up observations requires
a high sky density of sources spread in $cz$ but at similar declination.
Possible detections of low S/N will be observed in follow-up corroborating 
runs, which will help us quantify the reliability,
i.e. the fraction of candidate sources which observations
will confirm as real, as a function of the measured S/N. Those observations
will also deliver a significant addition of reliable detections, at a
very low ``cost'' of telescope time per source (see Papers I and II).

A few details should be kept in mind in the use of the contents of Table 1,
as described below:
\begin{itemize}
\item Several objects straddle or are confused by the HI emission from 
the Milky Way. A spectrum of the object is generally obtained by subtracting
a median spectrum obtained over a square perimetral segment of pixels around
the source, from the inner part of the perimeter populated by the source.
In some cases, interpolation of the spectrum across the Milky Way emission
provides an adequate result. Sources for which these intrusive techniques 
are applied are noted in the footnotes to Table 1.
\item Sources are often detected in the spectral vicinity of rfi features. 
Interpolation across the region affected by rfi is often necessary. While
the detection of the source may be clear, accurate measurement of its
parameters may be impossible and much caution should be adopted in their use. 
This problem is frequent for sources between
$cz\sim 8500$ \kms ~and $cz\sim 8900$ \kms, a spectral region not far removed
from the median redshift of our catalog. Each of these cases is noted in the footnotes
to Table 1.
\item Extended sources containing multiple concentrations, obviously physically 
associated, are sometimes found. We adopt the practice of measuring 
separately each of the concentrations and report them in Table 1 as separate
entries. They are also noted in the footnotes to Table 1.
\end{itemize}

Table 1 contains the principal parameters of HI candidate detections, namely:

Col. 1: an entry number for this catalog.

Col. 2: the source number in the Arecibo General Catalog, a private database
        of extragalactic objects maintained by M.P.H. and R.G. The AGC entry normally
        corresponds to the optical counterpart except in the cases of HI sources 
        which cannot be associated with an optical object with any high
        degree of probability.

Col. 3: center (J2000) of the HI source, after correction for systematic 
	telescope pointing errors, which range between a few and about a dozen 
	arcsec, as described in Section \ref{pos}.
	The accuracy of the HI positions depends on source strength. On average,
	the positional accuracy is about 24\arcsec. See Section \ref{pos} for
	details.

Col. 4: center (J2000) of the optical galaxy found to provide a reasonable
	optical counterpart. This position has been checked for each listed
	object and assessed using tools provided through the SDSS and {\it Skyview} websites.
	Quality of centroids is estimated to be 2\arcsec~ or better. The assessment
	of identification between optical and HI sources is based on spatial proximity, 
	redshift (if optical is available), morphology, color and evidence for optical
	emission lines (if known). For sources with no discernible optical
	counterpart and those for which such assignment is ambiguous, due to 
	the presence of more than one equally
	possible optical counterpart, no optical position is listed. The
	latter set includes HVCs. For objects with more than
	one possible candidate as an optical counterpart but such that one
	of the candidates is significantly more preferable than the others,
	an optical identification is made; however, a comment on
	the possible ambiguity is added in the notes to this table, as
	alerted by an asterisk in Col. 12.

Col. 5: heliocentric velocity of the HI source, $cz_{\odot}$, measured 
	as the midpoint 
	between the channels at which the flux density drops to 50\% of 
	each of the two peaks (or of one, if only one is present) at each
	side of the spectral feature. Units are \kms. The error on $cz_\odot$
	to be adopted is half the error on the width, tabulated in Col. 6.

Col. 6: velocity width of the source line profile, $W50$, measured at the 50\%
	level of each of the two peaks, as described for Col. 5. This value 
	is corrected for instrumental broadening. No corrections due to
	turbulent motions, disk inclination or cosmological effects are
	applied. Between brackets we show the
        estimated error on the velocity width, $\epsilon_w$, in \kms.
	This error is the sum in quadrature of two components: the first is a
	statistical error, principally dependent on the S/N
	ratio of the feature measured; the second is a systematic error
	associated with the subjective guess with which the observer estimates 
	the spectral boundaries of the feature: maximum and minimum guesses of 
	the spectral extent of the feature are flagged and the
	ratio of those values is used to estimate systematic errors on the
	width, the velocity and the flux integral. In the majority of cases,
	the systematic error is significantly smaller than the statistical
	error; thus the former is ignored.

Col. 7: integrated flux density of the source, $F_c$, in Jy \kms . This 
	is measured on the integrated spectrum, obtained by
	spatially integrating the source image over a solid angle of at
        least $7$\arcmin $\times 7$\arcmin ~and dividing by the sum of the survey beam
	values over the same set of image pixels (see Shostak \& Allen 1980). 
	Estimates of integrated fluxes for very extended sources with
	significant angular asymmetries can be misestimated by our 
	algorithm, which is optimized for measuring sources comparable with
	or smaller than the survey beam. A special catalog with parameters
	of extended sources will be produced after completion of the survey.
        The estimated uncertainty of the integrated flux density, in Jy \kms 
	is given between brackets. Uncertainties
	associated with the quality of the baseline fitting are not included;
	an analysis of that contribution to the error will be presented
	elsewhere for the full survey. See description of Col. 6 for the
	contribution of a possible systematic measurement error.

Col. 8: signal--to--noise ratio S/N of the detection, estimated as 
        \be
        S/N=\Bigl({1000 F_c\over W50}\Bigr){w^{1/2}_{smo}\over \sigma_{rms}}
        \ee
        where $F_c$ is the integrated flux density in Jy \kms, as listed in  col. 7,
        the ratio $1000 F_c/W50$ is the mean flux across the feature in mJy,
        $w_{smo}$ is either $W50/(2\times 10)$ for $W50<400$ \kms or
        $400/(2\times 10)=20$ for $W50 \geq 400$ \kms [$w_{smo}$ is a
        smoothing width expressed as the number of spectral resolution
        bins of 10 \kms bridging half of the signal width], and $\sigma_{rms}$
        is the r.m.s noise figure across the spectrum measured in mJy at 10
	\kms resolution, as tabulated in Col. 9. In a similar analysis, in 
        Giovanelli \etal ~2005b (hereafter Paper II) we adopted a maximum smoothing 
	width $W50/20=10$. See Figure \ref{himdist.ps} ~and related text
	below for details. The value of the smoothing width could be
	interpreted as an indication of the degree to which spectral smoothing 
	aids in the visual detection of broad signals, against broad--band 
	spectral instabilities. The ALFALFA
	data quality appears to warrant a more optimistic adoption of
	the smoothing width than previously anticipated. 

Col. 9: noise figure of the spatially integrated spectral profile, $\sigma_{rms}$,
	in mJy. The noise figure as tabulated is the r.m.s. and measured over the signal-- and
	rfi-free portions of the spectrum, after Hanning smoothing to a spectral
	resolution of 10 \kms.

Col. 10: adopted distance in Mpc, $D_{Mpc}$. For objects with $cz_{cmb}> 3000$, 
	the distance is simply  $cz_{cmb}/H_\circ$; $cz_{cmb}$ is the recession velocity
	measured in the Cosmic Microwave Background reference frame and $H_\circ$ is
	the Hubble constant,  for which we use a value of 70 \kms Mpc$^{-1}$.
	For objects of lower $cz_{cmb}$, we use a peculiar velocity model for the
	local Universe, as described in Paper II. Objects which are thought to be parts
	of clusters or groups are assigned the $cz_{cmb}$ of the cluster or group.
	Cluster and group membership are assigned following the method described
        in Springob \etal ~2006. A detailed
	analysis of group and cluster membership of ALFALFA objects will be presented
	elsewhere. Note that the Virgo cluster extends over much of the solid angle 
	surveyed. This introduces unavoidable ambiguities in the distance assignment,
	as the peculiar flow model only corrects for large--scale perturbations in the
	velocity field and is unable to deal effectively with regions in the
	immediate vicinity of the cluster and along a section of a cone which contains
	the cluster, up to $cz\sim 2500$ \kms. The distance to the Virgo cluster was
        assumed to be 16.7 Mpc.

Col. 11: logarithm in base 10 of the HI mass, in solar units. That parameter is 
	obtained by using the expression $M_{HI}=2.356\times 10^5 D_{Mpc}^2 F_c$. 

Col. 12: object code, defined as follows: 
	
	Code 1 refers to sources 
	of S/N and general qualities that make it a reliable detection.
	By ``general qualities'' we mean that, in addition to an approximate
	S/N threshold of 6.5, the signal should  exhibit a good match between
	the two independent polarizations and a spatial extent consistent
	with expectations given the telescope beam characteristics. Thus, some
	candidate detections with $S/N>6.5$ have been excluded on grounds
	of polarization mismatch, spectral vicinity to rfi features or peculiar
	spatial properties. Likewise, some features of $S/N<6.5$ are included
	as reliable detections, due to optimal overall characteristics of
	the feature. The S/N threshold for acceptance of a reliable detection
	candidate is thus soft. In a preliminary fashion, we estimate that
	detection candidates with $S/N>6.5$ in Table 1 are reliable, i.e. they
	will be confirmed in follow--up observations in better than 95\% of
	cases (Saintonge 2007). Follow-up observations planned for 2007 will 
	set this estimate on stronger statistical grounds.

	Code 2 refers to sources of low S/N ($<$ 6.5), which would  
        ordinarily not be considered
	reliable detections by the criteria set for code 1. However, those
	HI candidate sources are matched with optical counterparts with known
	optical redshifts which match those measured in the HI line. These
	candidate sources, albeit ``detected'' by our signal finding algorithm,
	are accepted as likely counterparts only because of the existence of
	previously available, corroborating optical spectroscopy. We refer to
	these sources as ``priors''. We include them in our catalog because
	they are very likely to be real.

	Code 9 refers to objects assumed to be HVCs; no
	estimate of their distances is made.

 	Notes flag. An asterisk in this column indicates a comment is included
	for this source in the text below.

Only the first few entries of Table 1 are listed in this version of the
paper. The full contents of Table 1 are accessible through the ALFALFA website
at {\it http://egg.astro.cornell.edu/alfalfa}.

%[NOTE: THE FULL TABLE IS INCLUDED HERE FOR REFEREE ACCESS, BUT ONLY THE FIRST PAGE IS TO
%APPEAR IN THE PAPER JOURNAL.]

\begin{deluxetable}{crccrcrrrrrl}
\rotate
\tablewidth{0pt}
\tabletypesize{\scriptsize}
\tablecaption{HI Candidate Detections\label{alfadet}}
\tablehead{
\colhead {Cat. nr} & {AGC}   & \colhead{HI Coords } & \colhead{Opt. Coords.} &
\colhead{cz$_\odot$}  & \colhead{$w50 ~(\epsilon_w$)} &
\colhead{$F_{c} ~(\epsilon_{f})$} & \colhead{S/N} & \colhead{rms} &  
\colhead{Dist}    & \colhead{$\log M_{HI}$} & \colhead{Code} 
    \\
 & & (J2000) & (J2000) & {\kms} & {\kms} & {Jy \kms} & & {mJy} & Mpc & {$M_\odot$} &  
}
\startdata
1-  1  & 210692 & 114419.8+150620 & 114420.2+150616 &  10489 &   283( 13) &   1.42(0.08) &    8.6 &   1.85 &  151.5 &   9.89 & 1   \\
1-  2  & 215418 & 114421.4+150459 &                 &    247 &    33(  2) &   0.38(0.04) &    6.8 &   2.09 &        &        & 9 * \\
1-  3  & 213528 & 114432.1+131527 & 114432.0+131511 &  10291 &   119( 12) &   0.52(0.06) &    4.8 &   2.09 &  148.6 &   9.43 & 2   \\
1-  4  & 213473 & 114437.1+124657 & 114439.5+124726 &  12751 &   245( 88) &   1.06(0.09) &    5.8 &   2.14 &  183.7 &   9.93 & 2   \\
1-  5  & 212928 & 114440.6+144551 & 114440.9+144555 &  10240 &   379(  4) &   2.32(0.08) &   14.0 &   1.73 &  147.9 &  10.08 & 1   \\
1-  6  & 215197 & 114442.2+150215 & 114444.0+150140 &   3354 &   103( 40) &   0.72(0.05) &    7.1 &   1.78 &   50.4 &   8.63 & 1   \\
1-  7  & 215148 & 114443.7+121708 & 114442.6+121802 &  10258 &   303( 76) &   1.20(0.09) &    6.5 &   2.11 &  148.1 &   9.79 & 1   \\
1-  8  & 210746 & 114524.6+142242 & 114522.1+142200 &   3342 &   129( 25) &   0.90(0.05) &   10.8 &   1.56 &   50.3 &   8.73 & 1   \\
1-  9  & 210753 & 114533.5+121240 & 114534.9+121218 &   9298 &    92(  3) &   1.40(0.06) &   14.4 &   2.05 &  134.5 &   9.78 & 1   \\
1- 10  & 215149 & 114554.6+134955 & 114556.1+135021 &   3267 &    38(  9) &   0.57(0.04) &    7.6 &   1.77 &   49.2 &   8.51 & 1   \\
1- 11  & 211216 & 114609.2+125231 & 114609.0+125246 &   3296 &   145(  6) &   1.45(0.05) &   15.9 &   1.58 &   49.6 &   8.92 & 1   \\
1- 12  &   6747 & 114622.9+134937 & 114624.1+134938 &   2696 &    95(  4) &   6.14(0.06) &   58.7 &   1.82 &   41.2 &   9.39 & 1   \\
1- 13  &   6753 & 114646.5+143208 & 114646.2+143158 &   3149 &   169(  6) &   2.10(0.06) &   18.2 &   1.73 &   47.5 &   9.05 & 1   \\
1- 14  & 210779 & 114700.3+135223 & 114659.7+135225 &   3165 &   231(  1) &   1.97(0.06) &   15.9 &   1.63 &   47.2 &   9.01 & 1   \\
1- 15  &   6758 & 114706.2+134238 & 114706.3+134223 &   3103 &   178(  2) &   7.15(0.06) &   67.2 &   1.63 &   47.2 &   9.57 & 1   \\
1- 16  & 215618 & 114720.4+152015 & 114716.8+151924 &   6401 &   162( 12) &   0.57(0.07) &    4.6 &   2.18 &   93.4 &   9.07 & 2   \\
1- 17  & 213478 & 114725.7+121308 & 114725.4+121239 &   6111 &   212( 48) &   0.64(0.08) &    4.5 &   2.18 &   89.3 &   9.08 & 2   \\
1- 18  & 215151 & 114750.3+134144 & 114750.7+134215 &   3456 &    72(  4) &   0.61(0.05) &    7.3 &   1.79 &   51.9 &   8.59 & 1   \\
1- 19  & 213479 & 114800.7+120714 & 114800.2+120649 &   9218 &   102(  8) &   0.85(0.06) &    7.2 &   2.16 &  133.3 &   9.55 & 1   \\
1- 20  & 215198 & 114801.1+145126 & 114800.4+145221 &  11840 &   336( 23) &   1.29(0.07) &    6.8 &   1.51 &  170.8 &   9.95 & 1 * \\
1- 21  & 210798 & 114809.0+125445 & 114808.8+125453 &   6216 &   226(  1) &   3.37(0.06) &   25.5 &   1.68 &   90.8 &   9.82 & 1   \\
1- 22  &   6775 & 114812.6+131252 & 114812.7+131230 &   3165 &   193(  0) &   7.89(0.07) &   59.1 &   1.99 &   47.1 &   9.62 & 1   \\
1- 23  & 212846 & 114816.9+124330 & 114817.9+124333 &   3955 &   191( 19) &   3.82(0.08) &   31.1 &   1.98 &   58.9 &   9.49 & 1 * \\
1- 24  & 210799 & 114821.6+124321 & 114820.1+124300 &   3969 &   273( 17) &   4.51(0.09) &   24.1 &   1.99 &   59.1 &   9.57 & 1 * \\
1- 25  & 210804 & 114830.6+124401 & 114830.7+124347 &   3629 &   157( 10) &   1.72(0.08) &   12.3 &   2.03 &   54.3 &   9.08 & 1 * \\
1- 26  & 210807 & 114843.8+140323 & 114844.0+140310 &   3206 &   224(  8) &   1.53(0.07) &   10.9 &   1.85 &   47.2 &   8.90 & 1   \\
1- 27  & 210809 & 114858.1+155322 & 114856.0+155325 &  17096 &   399(  4) &   1.88(0.10) &   69.7 &   2.03 &  245.8 &  10.43 & 1   \\
1- 28  & 215154 & 114901.2+133708 & 114904.4+133746 &   2981 &    33( 10) &   0.51(0.03) &    8.7 &   1.47 &   45.1 &   8.39 & 1 * \\
1- 29  & 215156 & 114901.9+134038 & 114903.1+134052 &   7136 &   205( 27) &   1.04(0.06) &    8.8 &   1.71 &  103.8 &   9.42 & 1   \\
1- 30  & 213484 & 114908.0+123721 & 114912.2+123754 &   4054 &   206( 35) &   0.73(0.08) &    4.7 &   2.16 &   60.3 &   8.80 & 2   \\
1- 31  & 215200 & 114913.0+153708 & 114914.1+153651 &  16951 &   376( 29) &   2.13(0.11) &   10.1 &   2.31 &  243.7 &  10.47 & 1   \\
1- 32  & 210814 & 114920.7+152426 & 114920.1+152432 &   6371 &   163(  2) &   1.66(0.07) &   12.2 &   2.27 &   93.0 &   9.53 & 1   \\
1- 33  & 215158 & 114942.0+122404 & 114940.1+122338 &   3201 &   129(  4) &   0.73(0.05) &    8.0 &   1.70 &   48.2 &   8.60 & 1   \\
1- 34  & 210820 & 114943.4+131455 & 114941.4+131441 &   6141 &   207(  2) &   1.29(0.07) &    9.8 &   1.82 &   89.7 &   9.39 & 1   \\
1- 35  & 210822 & 115002.6+150132 & 115002.7+150124 &    756 &    45(  3) &   1.49(0.04) &   18.2 &   1.70 &    8.6 &   7.42 & 1   \\
1- 36  & 215229 & 115007.5+154613 & 115008.0+154702 &  13303 &   338( 86) &   1.76(0.10) &    9.6 &   2.09 &  191.7 &  10.18 & 1 * \\
1- 37  & 215202 & 115011.4+143926 & 115009.7+143918 &   6485 &   192(  4) &   1.39(0.08) &    9.7 &   2.08 &   94.6 &   9.47 & 1   \\
1- 38  & 213488 & 115018.9+124333 & 115016.4+124327 &   7774 &   203(  5) &   0.88(0.08) &    5.5 &   2.28 &  112.8 &   9.42 & 2   \\
1- 39  & 215203 & 115038.1+142719 & 115037.8+142712 &   6082 &    76(  3) &   0.72(0.04) &    9.1 &   1.63 &   88.9 &   9.13 & 1   \\
1- 40  & 215204 & 115045.4+144538 & 115044.8+144633 &  16443 &   311( 38) &   1.07(0.08) &    5.5 &   1.88 &  236.5 &  10.15 & 2   \\
\hline
\enddata
\end{deluxetable}

\subsection{Notes to Tabulated Data}\label{notes}
	
Notes associated with the objects listed in Table 1 follow. Each note is preceded by
the catalog entry name as listed in Col. 1 of Table 1.

{\footnotesize
%\noindent HI114347.2+121508	other possible oc at 114350.2+121455 and 114345.8+121439

%\noindent HI114358.6+132104	other possible counterpart candidate at 114407.0+132027

\noindent 1-2:		extended HVC?

\noindent 1-20:		also possible opt id with companion at 114802.6+145228

\noindent 1-21:		blend w/210799, params. very uncertain

\noindent 1-22:		blend w/212846, params. very uncertain

\noindent 1-23:		emission of 210799 also in spectrum

\noindent 1-28:		optical identification uncertain 114904.4+133746; also possible 114858.8+133710, 
			nearer to HI but $\sim$1 mag fainter

\noindent 1-36:		possible blend; alternative optical identification is 115012.8+154620, opt z=0.045,
			which is 1 mag brighter, but twice as far from HI center

\noindent 1-42:		extended HI, blended with emission to NE

\noindent 1-44:		no identifiable optical counterpart: appendage of 215231?

\noindent 1-46:		HI emission merges in spectral region w/strong rfi; HI parms.
			very uncertain

\noindent 1-62:		no clear opt counterpart; UGC 6911 5.3\arcmin ~to W, similar $cz$;
			HI feature appears real, not sidelobe effect

\noindent 1-70:		HI emission on edge of bandpass, poor sensitivity,
			params uncertain

\noindent 1-98:		very near edge of bandpass: poor sensitivity, params uncertain

\noindent 1-109:	alternative opt id: 120521.3+153110, a much fainter obj but
			nearer to HI position

\noindent 1-113:	120625.3+132303 is in a pair with 120626.0+132254; HI emission
			could be associated with either or both

\noindent 1-116:	no opt counterpart, extended HI: compact HVC (cHVC)?

\noindent 1-122:	no opt counterpart, extended HI: HVC

\noindent 1-124:	HI emission on edge of bandpass, poor sensitivity,
			parms uncertain

\noindent 1-129:	HI merges in spectral region affected by rfi;
			parms of detection very uncertain

\noindent 1-130:	no optical counterpart; extended HI: HVC projected
			in vicinity of NGC 4192

\noindent 1-148:	no optical counterpart; extended HI: HVC projected
			in vicinity of NGC 4192

\noindent 1-153:	no optical counterpart; extended HI: HVC projected
			in vicinity of NGC 4192

\noindent 1-157:	no optical counterpart; extended HI: HVC projected
			in vicinity of NGC 4192

\noindent 1-158:	HI emission may have SW to NE extension

\noindent 1-162:	no optical counterpart; extended HI: HVC projected
			in vicinity of NGC 4192

\noindent 1-173:	no identifiable optical counterpart

\noindent 1-184:	no identifiable optical counterpart; extended HI: HVC projected
			in vicinity of NGC 4192

\noindent 1-185:	no identifiable optical counterpart

\noindent 1-210: 	no optical counterpart, S of U7284" galaxy appendage or HVC?

\noindent 1-214:	HI pos matches that of IC 3080, which has discordant opt z; match
			with 121607.0+141237 based on similarity of z, but note large
			pos discrepancy: ambiguous opt identification

\noindent 1-217:	no identifiable optical counterpart; extended HI: HVC projected
			in vicinity of NGC 4192

\noindent 1-228:	no  identifiable optical counterpart; extended HI: HVC projected
			in vicinity of NGC 4192

\noindent 1-229:	alternative optical identification is 121716.7+142732, opt brighter but farther
			from HI center 

\noindent 1-235:	no identifiable optical counterpart; portion of Virgo~HI21 (Davies \etal ~2004;
                        Minchin \etal ~2005)

\noindent 1-238:	no identifiable optical counterpart; portion of Virgo~HI21 (Davies \etal ~2004;
                        Minchin \etal ~2005)

\noindent 1-239:	no identifiable optical counterpart; portion of Virgo~HI21 (Davies \etal ~2004;
                        Minchin \etal ~2005)

\noindent 1-241:	in vicinity of Virgo~HI21 (Davies \etal ~2004;
                        Minchin \etal ~2005)
		
\noindent 1-242:	no identifiable optical counterpart; unresolved, compact HVC projected
			in vicinity of NGC 4192

\noindent 1-245:	no identifiable optical counterpart; unresolved, compact HVC projected
			in vicinity of NGC 4192

\noindent 1-257:	no identifiable optical counterpart; 4\arcmin ~N of HI121910.9+125322=A220351:
			tidal appendage?

\noindent 1-260:	resolved disk, elongated SW (hi vel) to NE (lo vel);
			previous HI cz=1280 appears to be wrong

\noindent 1-263:	crowded opt field, ambiguous opt id; alternative opt counterparts
			are 121942.5+132549, 121948.4+132522, 121945.1+132627

\noindent 1-264:	no identifiable optical counterpart; barely resolved, compact HVC

\noindent 1-271:	no identifiable optical counterpart; VCC429 at 122043.8+143751 and
			similar z is at 2.3\arcmin ~to NE; HI assumed related.

\noindent 1-276:	blend of U7412 and U7418 (N4298 and N 4302); HI flux mainly of U7418

\noindent 1-278:	opt counterpart is close galaxy pair

\noindent 1-279:	no identifiable optical counterpart; unresolved, compact HVC

\noindent 1-287:	v uncertain separation from MW HI; marginal detection and v poor HI parms

\noindent 1-309:	uncertain separation from MW HI

\noindent 1-312:	v poor positional match, marginal S/N, id

\noindent 1-316:	3.6\arcmin ~to SE of M86=U7532, no identifiable opt counterpart; part of 
                                Virgo~HI4 (Davies \etal ~2004; note that the declination listed in that
                                paper is incorrect) and shown to be a plume extending from NGC~4388,
                                by Oosterloo \& van Gorkom (2005)

\noindent 1-318:	uncertain separation from MW HI

%\noindent HI122711.5+140754	no identifiable optical counterpart; marginal S/N
%
\noindent 1-336:	extended HI source, blend of 123115.0+141148 and 123120.0+131144;
			id assignment of HI to 123115.0+141148 on vicinity grounds

\noindent 1-347:	fainter opt galaxy at 123355.2+135554 also possible counterpart

\noindent 1-352:	no clearly identifiable opt counterpart

\noindent 1-354:	no identifiable opt counterpart; in spite of fair S/N, doubts
			on reality, due to standing waves produced by vicinity to M87

\noindent 1-359:	opt counterpart is extremely lsb, v extended object?
			however, vicinity of M87 makes HI detection somewhat doubtful

\noindent 1-360:	no identifiable opt counterpart, marginal S/N

\noindent 1-366:	no identifiable opt counterpart

\noindent 1-371:	ambiguous opt id: 123650.8+141506 possible;
			224865, identified with 123643.1+141611, 2.7\arcmin ~to NW, is
			at similar z (NGC 4571 in foreground)

\noindent 1-377:	optical id ambiguous: other possibilities are 123857.5+142435
			and 123859.1+142457

\noindent 1-386:	on edge of band, ragged data

\noindent 1-391:	blend with 7874=N4633, interact syst, 
			and interference with MW HI; parms uncertain

\noindent 1-392:	most of emission in profile is associated with source
			U7874, identified with U7884=N4639, 2.8\arcmin ~E;
			feature measured is wing to high velocity side of emission peak

\noindent 1-397:	extended emission of U7902=N4654 overwhelms
			the field; emission tentatively assigned to opt galaxy at
			124412.0+125631 is low velocity wing of line

\noindent 1-398:	positional offset of 1\arcmin ~between opt and HI probably real,
			not centroiding error

\noindent 1-399:	optical identification with very faint lsb feature, 3.5\arcmin ~to SE of IC3720

\noindent 1-402:	ambiguous opt id: also possible 124514.7+141906

\noindent 1-409:	on edge of band, ragged data

\noindent 1-417:	optical identification with faint blue obj is very tentative

\noindent 1-426:	other possible opt counterpart at 125209.6+150456, marginally farther away

\noindent 1-451:	alternative opt id: fainter obj at 125911.0+142519

\noindent 1-474:	optical identification with bluest galaxy in triplet

\noindent 1-488:	on edge of band, ragged data

\noindent 1-493:	on edge of band, ragged data

\noindent 1-494:	no identifiable opt counterpart, extended HI: 
			part of HVC complex

\noindent 1-497:	no identifiable opt counterpart, extended HI: 
			part of HVC complex

%\noindent HI131056.8+115404	on edge of spectral region affected by rfi; parms uncertain

\noindent 1-500:	no identifiable opt counterpart, extended HI: 
			part of HVC complex

\noindent 1-502:	no identifiable opt counterpart, extended HI: 
			part of HVC complex

\noindent 1-504:	gal emission merges in region affected by rfi: parms uncertain

\noindent 1-507:	gal emission merges in region affected by rfi: parms uncertain

\noindent 1-513:	no identifiable opt counterpart, extended HI: 
			part of HVC complex

\noindent 1-524:	gal emission merges in region affected by rfi: parms uncertain

\noindent 1-531:	on edge of region affected by rfi; id with very faint, lsb 
			object is tentative; caveat emptor

\noindent 1-532:	gal emission merges in region affected by rfi: parms uncertain

\noindent 1-534:	on edge of band, ragged data

\noindent 1-538:	gal emission merges in region affected by rfi: parms uncertain

\noindent 1-541:	gal emission merges in region affected by rfi: parms uncertain

\noindent 1-543:	gal emission merges in region affected by rfi: parms very uncertain

\noindent 1-556:	no identifiable opt counterpart: compact HVC

\noindent 1-563:	gal emission merges in region affected by rfi, but params only
			mildly affected

%\noindent HI132429.8+134812	no unambiguous opt counterpart; possibilities are 132426.3+134859,
%			132429.7+134758 (v faint); HI132425.0+134854 in the foreground

\noindent 1-578:	no unambiguous opt counterpart; blue obj in vicinity; marginal detection

\noindent 1-611:	N5221, highly disturbed

\noindent 1-613:	optical identification assigned to blue obj superposed onto E galaxy at similar {\it cz}:
			merger underway? N5221 is 6\arcmin ~N, highly disturbed

\noindent 1-621:	on edge of band, ragged data

\noindent 1-636:	ambiguous opt id: possible opt counterpart are 134124.7+151630 and 134119.4+151553

\noindent 1-639:	HI source may be extended to NW and SE of center

\noindent 1-645:	ambiguous opt id; 134233.9+130210 is alternative candidate, 
			20\arcsec ~farther from HI center

\noindent 1-665:	marginal det; tight pair;  other possible opt counterpart at 134739.2+154404

\noindent 1-667:	blend; optical counterpart is interacting pair VII Zw 338

\noindent 1-673:	ambiguous opt id; 134941.6+155702  is alternative candidate 

\noindent 1-692:	extended HI, no opt counterpart: IVC

\noindent 1-700:	HI emission merges in rfi; params. highly uncertain; opt gal in pair,
				alternative possible opt. counterpart at 135452.3+140741

\noindent 1-710:	emission blended with 233714, parms uncertain

\noindent 1-711:	see 230859

\noindent 1-722:	blend with emission by several other objects within 1.5\arcmin

%\noindent HI140020.7+125742	blend of emission by 140020.7+125742 and 140025.3+125728=U8920

}

%FIGURE 2
\begin{figure}[h]
%\figurenum{1}
\plotone{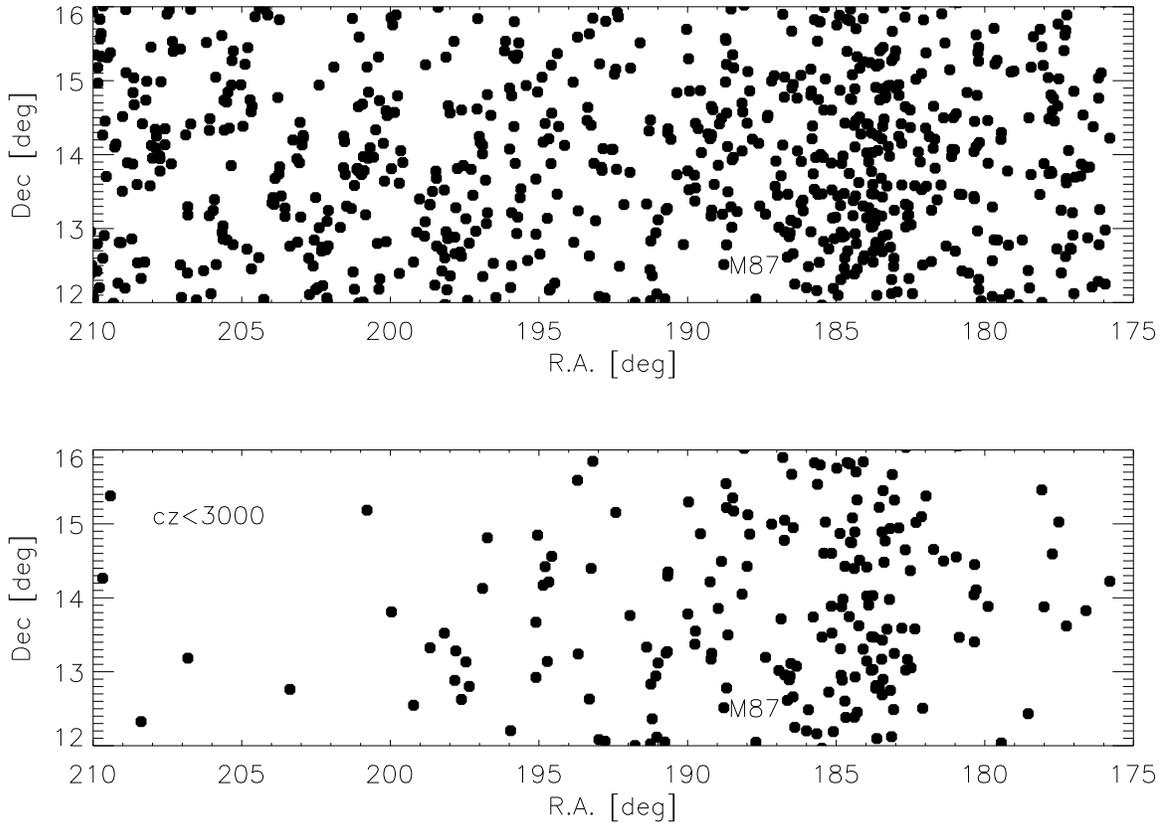}
\caption{Sky distribution of HI candidate detections listed in Table 1:
all objects in upper panel, objects with $cz<3000$ \kms ~ in the lower
one. A label identifies the position of M87, in the vicinity of which
HI sources cannot be effectively surveyed because of standing waves
and large increase in system temperature contributed by its strong
continuum emission; see Section \ref{m87}.}
\label{radec.ps}
\end{figure}

\section{Statistical Properties of the ALFALFA HI Sample}\label{stats}

The distribution by detection code of HI detections listed in Table 1 is as follows:
\nhvc ~(4\%) are HVCs, \nprior ~(16\%) are of type ``2'', i.e. 
``priors'' of low S/N but corroborated by optical redshifts, and \ndet ~($\sim$80\%) 
are detections of type "1", of quality fair to excellent.

The 132 \sqd ~region of the sky covered by the sample of HI detections listed in Table 1 
has been intensely investigated by a variety of studies, including surveys of
the Virgo cluster and the SDSS. Targeted observations of optically selected
samples amounting to many hundreds of hours of telescope time have yielded
detections for 158 objects in the {\it GOLDMINE} compilation of
Gavazzi \etal ~(2003), some of which are fainter than the ALFALFA detection
limit (and required a lot more integration time).
The blind HI Virgo HIJASS survey (Davies \etal ~2004)
is partly contained within the survey area presented here. In the HIJASS region
of full sensitivity which overlaps our ALFALFA dataset, HIJASS found 15 HI
detections; in contrast, we detect 193 HI sources in the same region. We detect
both the HI plume found near NGC~4388 (Oosterloo \& van Gorkom 2005) referred
to as Virgo~HI4  by Davies \etal ~2004 and the extended emission
in the vicinity of the ``dark galaxy'' Virgo~HI21 (Davies \etal ~2004;
Minchin \etal ~2005) showing a clear connection to NGC~4254; we will discuss
this structure elsewhere (Haynes \etal ~in preparation).
Likewise, the improvement over HIPASS is impressive. In the region containing the \ncat
~HI detections included in Table 1, the HIPASS dataset includes only 40 objects
(Wong \etal ~2006). Of those, several show large position offsets with respect 
to multiple ALFALFA detections lying within a single Parkes beam,
and two are not confirmed by ALFALFA. 

The median distance of this sample is 102 Mpc.
As of late August 2006, 72\% of the ALFALFA
HI detections presented here have optical or HI redshifts previously known and
28\% are new. The percentage of ALFALFA sources which are new HI detections is 69\%.
In view of the fact that numerous HI studies have been conducted in this
region, largely based on optically selected targets, the percentage of 69\% {\it new}
HI detections illustrates the fact that previous HI surveys, generally based on
flux-- or size--limited, optically selected samples, missed the majority
of HI sources: the conventional wisdom on which optical targets would turn
out to be HI--rich appears to have been limited.

About 25\% of ALFALFA detections have $cz < 3000$ \kms. The fraction of
local objects in the catalog presented here is enhanced 
by the fact that the region sampled crosses the supergalactic plane and 
the northern part of the Virgo cluster, one of the densest regions of the
Local Supercluster. The detection rate, of about 5.4 objects (4.4 of detection 
code 1) per square degree, is thus enhanced by a circumstance associated
with the characteristics of the large--scale structure of the galaxy distribution
in the local Universe. Were it not for the enhancement in the detections within
the Local Supercluster, the detection rate would have been lower by about one 
fifth. This can be visually gauged by inspection of Figure \ref{radec.ps}, which
shows the sky distribution of all the detected sources in the upper panel and that 
of objects with $cz < 3000$ \kms ~in the lower panel.

%FIGURE 3
\begin{figure}[ht]
\epsscale{0.7}
%\figurenum{1}
\plotone{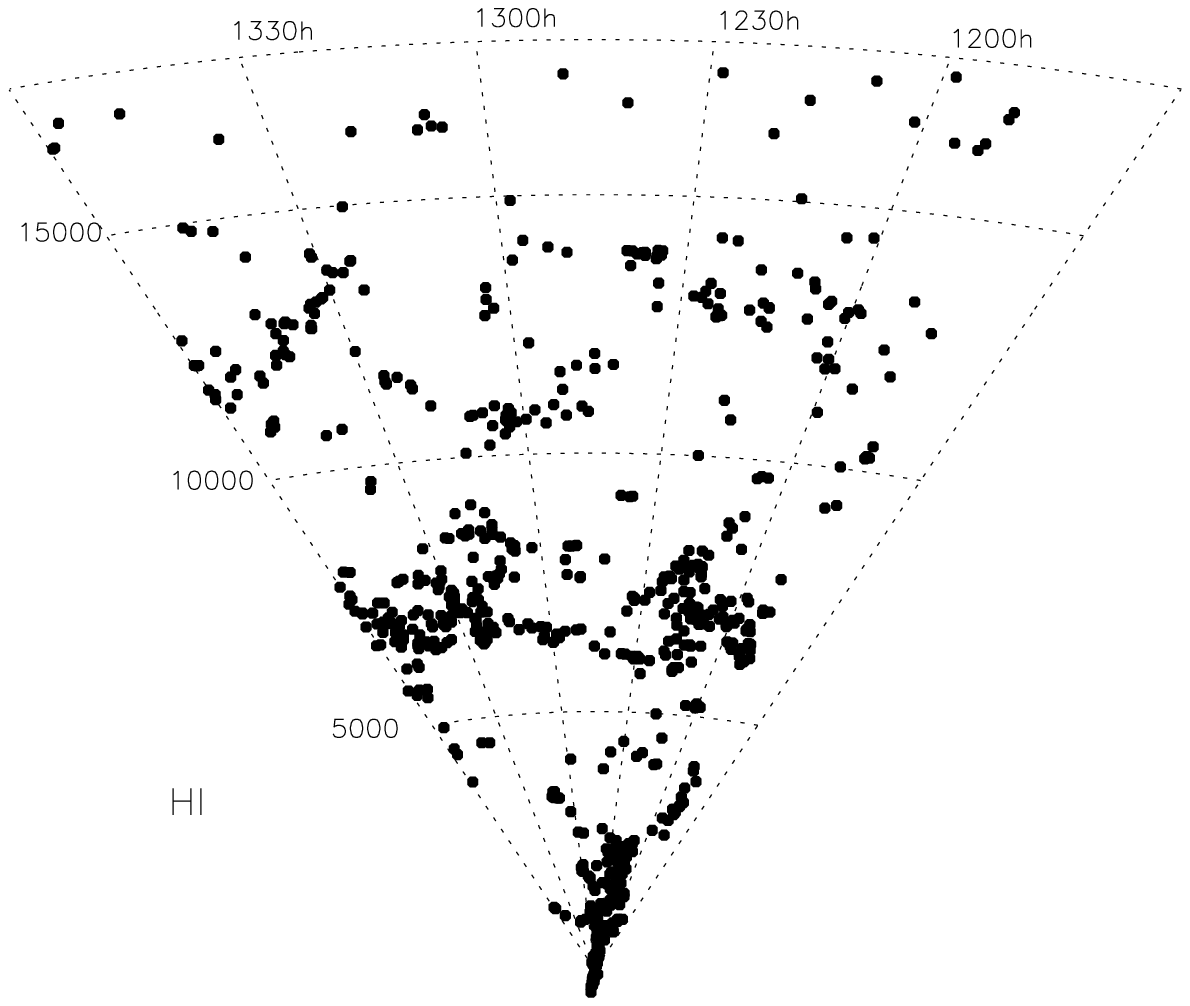}
\vskip -3cm
\plotone{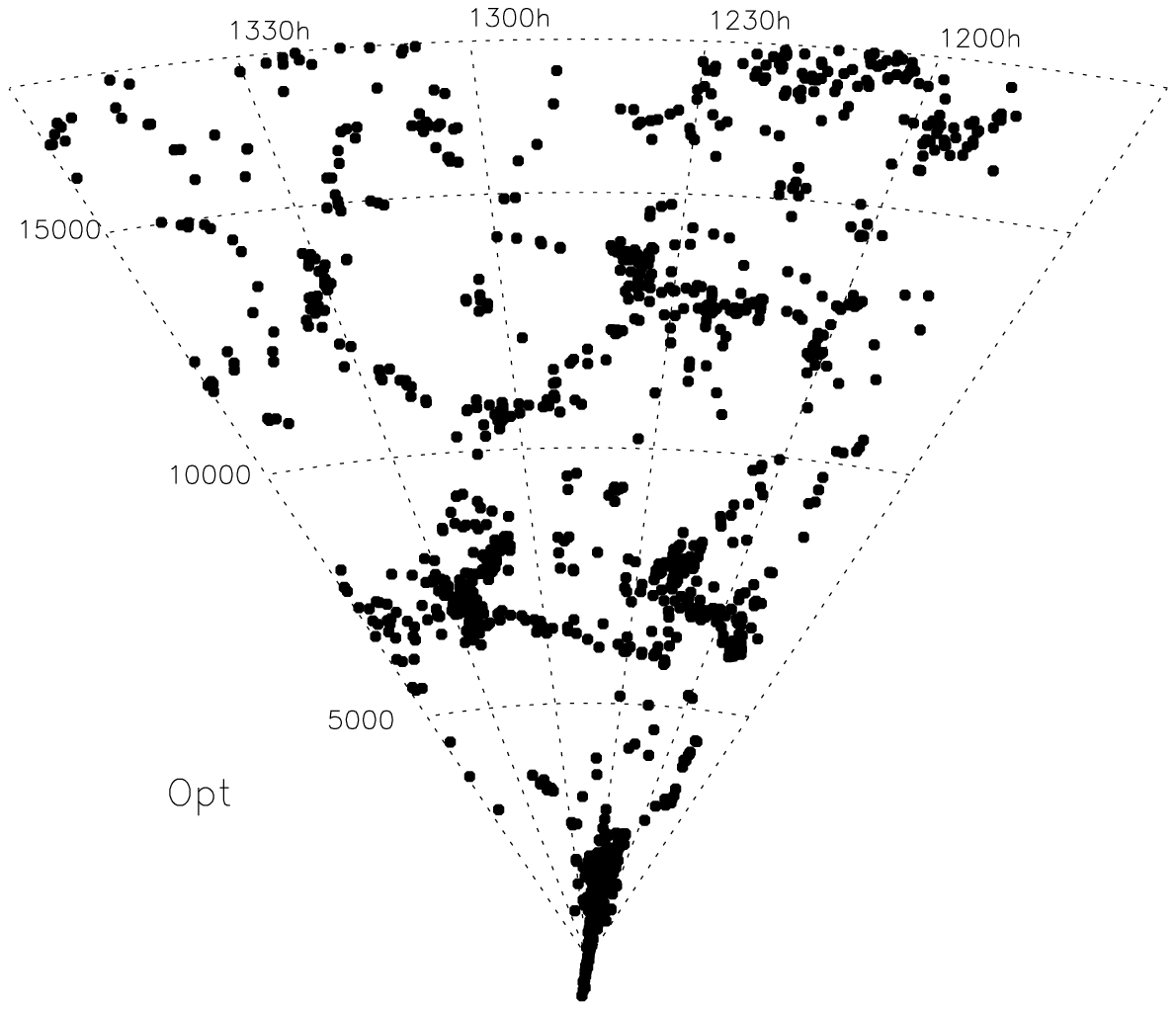}
\vskip -1cm
\caption{Right Ascension vs. recession velocity in \kms, of all sources
listed in Table 1 (upper panel ``HI'') and analogous plot of galaxies with
optical redshifts in the same region of sky (lower panel ``Opt''). Note the 
Virgo cluster in the lower portion of each cone plot.}
\label{cone.ps}
\end{figure}

As mentioned in Section \ref{m87}, the region of about $1^\circ$ to 1.5$^\circ$ radius 
about the location of M87 is contaminated by the high brightness of its
strong radio source which not only increases the system temperature but
also induces strong standing waves. Therefore, the sky region around M87
cannot be sampled by HI spectroscopy, except for the very brightest HI sources.
The location of M87 is indicated in 
Figure \ref{radec.ps}, at the center of a region of rarefied density of
HI detections.

%FIGURE 4
\begin{figure}[ht!]
%\figurenum{1}
\epsscale{0.5}
\plotone{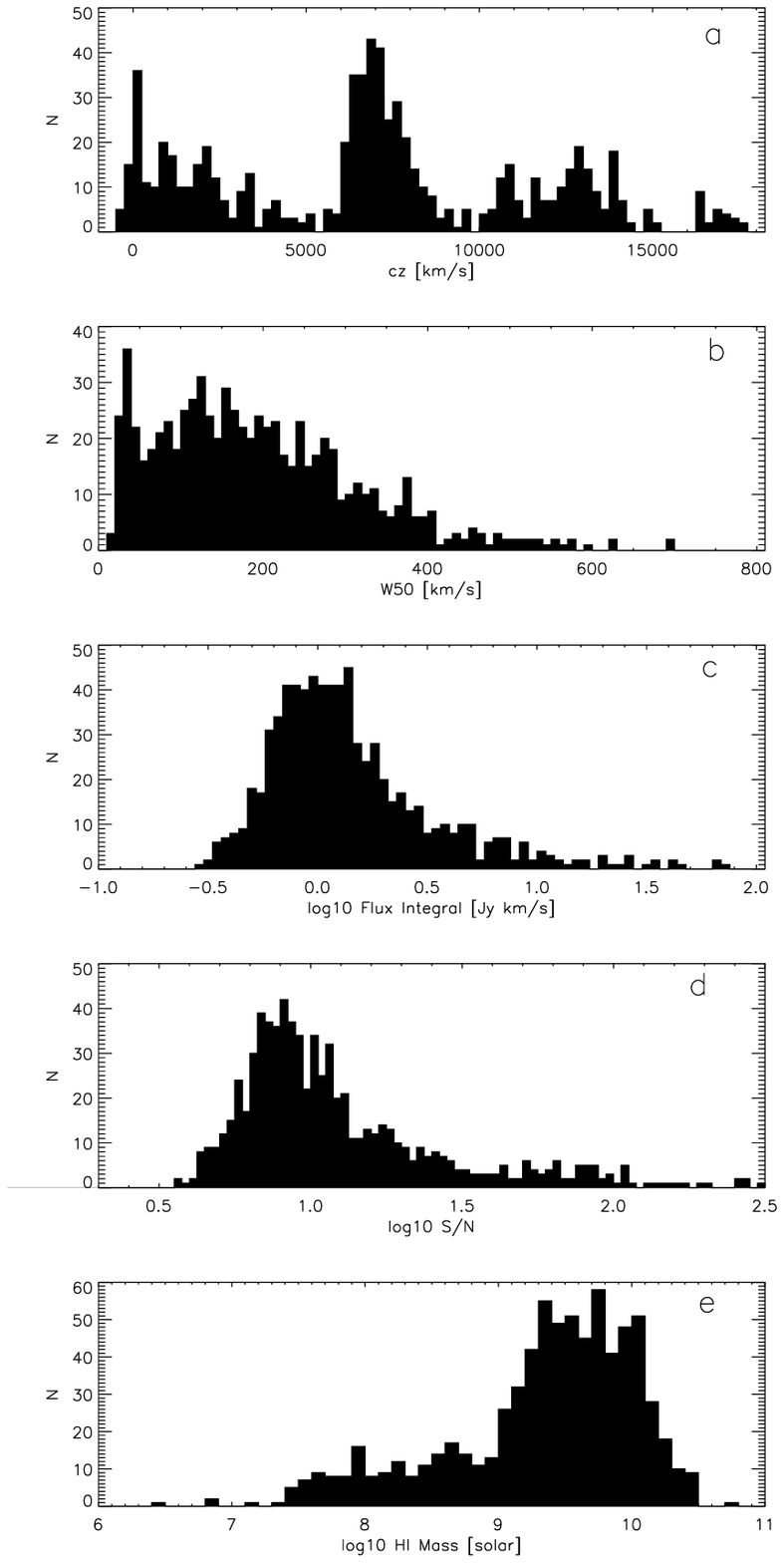}
\caption{Histograms of the HI candidate detections listed in Table 1.
From top to bottom: (a) recessional velocity in \kms; (b) HI line width W$_{50}$ in \kms;
(c) log base 10 of the flux integral in Jy 
\kms; (d) the peak signal-to-noise
ratio, in logarithmic units; (e) the derived HI mass, in logarithmic
units of \msun. Note the ``step'' in the HI mass histogram near $10^{7.5}$,
which is probably an artifact due to uncertainties in galaxy distances in the
direction towards the Virgo cluster, as decribed in the text.}
\label{histos.ps}
\end{figure}

The upper panel of Figure \ref{cone.ps} shows the redshift distribution of the 
galaxies in the current sample. The signatures of the Virgo cluster at the low 
velocity end and of other features in the large--scale structure of the local
Universe are clearly evident. The lack of detections on a strip nearly 1000 \kms
~wide near $cz\simeq 15000$ \kms reflects a bias of the survey. It corresponds
to a spectral region heavily and nearly continuously affected by rfi, as shown
in Figure \ref{weights.ps}. The lower panel of Figure \ref{cone.ps} shows the
analogous diagram for the galaxies with known optical redshifts in the same region
of the sky. The large-scale distribution properties of both samples are very
similar. The most notable difference between the two plots is the rarefaction of
sources at higher $cz$ in the HI panel.

%FIGURE 5
\begin{figure}[h!]
%\figurenum{1}
\epsscale{0.7}
\plotone{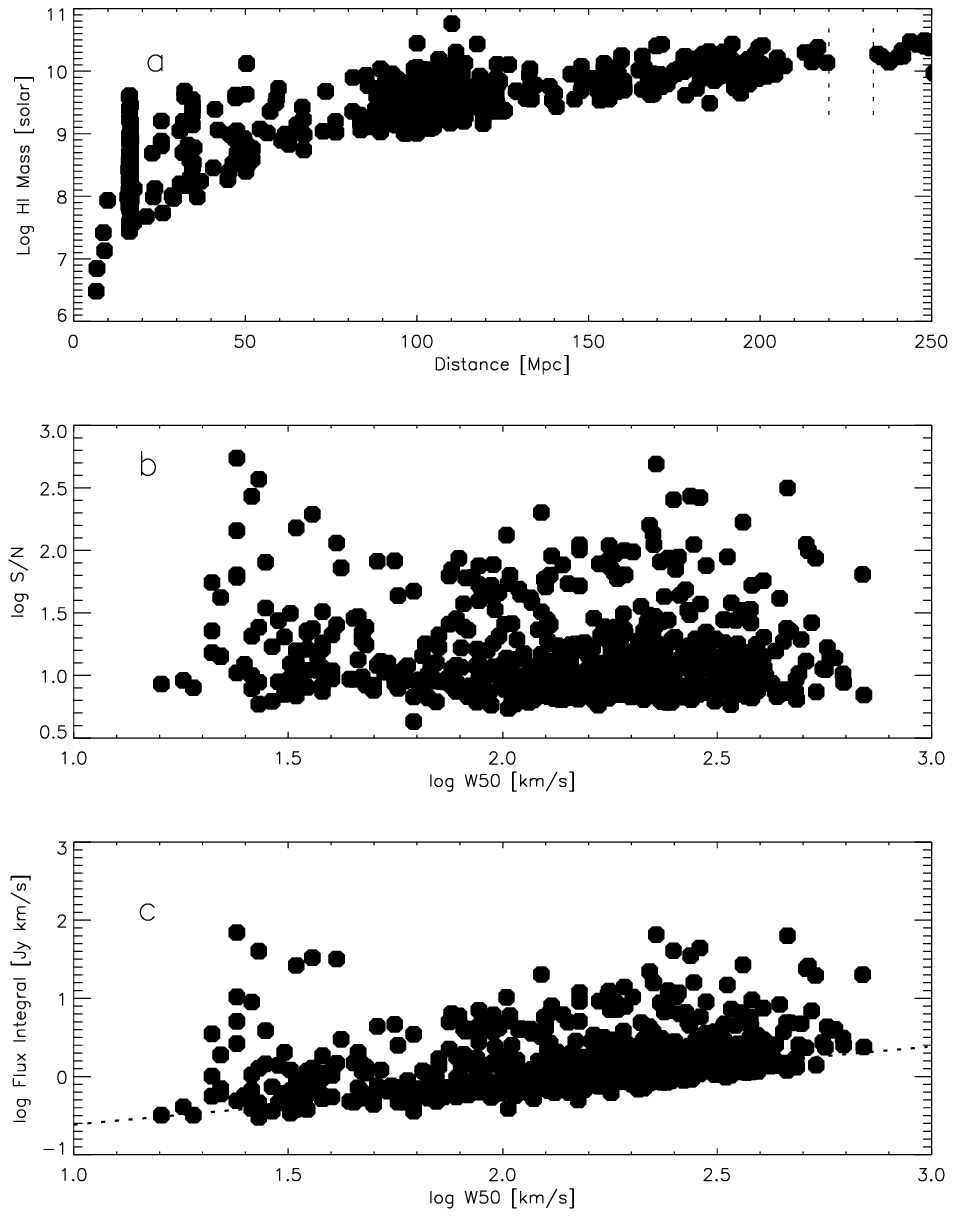}
\caption{(a) HI mass vs. distance for all sources of types 1 and 2
in Table 1. The Virgo cluster is the vertical feature near D=17 Mpc.
Cosmic sources within the spectral region between the two vertical 
dashed lines are inaccessible due to rfi. 
(b) S/N vs. velocity width for all sources in Table 1. The lower envelope
appears to be independent of S/N, indicating that no significant bias
is present in the detection of sources of large width.
(c) Flux integral vs. velocity width. The dashed line corresponds
to a S/N limit of 6.5.}
\label{himdist.ps}
\end{figure}

Figure \ref{histos.ps} ~summarizes the distribution in the values of 
heliocentric velocity $cz_\odot$, velocity width $W50$, flux integral $F_c$,
signal--to--noise ratio S/N and HI mass $M_{HI}$ for the candidate HI 
detections reported in this catalog. We remind the reader that this survey
is not designed to be complete to a given flux integral limit, but rather
that the sensitivity limit is a function of velocity width, as discussed
in Paper I, Paper II and further below. We also note the paucity of sources with
reported HI mass below $10^7$ $M_\odot$ in Figure \ref{histos.ps}(e); 
while this may be due to a truly low cosmological population of low mass 
HI sources, the effect is likely to be, at least in part, spurious. As distances have been inferred
from redshifts alone, albeit with the aid of a peculiar velocity flow model,
the model yields highly uncertain distances within the region sampled, due
to the presence of the Virgo cluster: a very nearby, low velocity object 
is far more likely to be assigned the Virgo cluster distance by the flow
model than a Virgo cluster object of low velocity to be assigned a
low distance. The presence of the Virgo cluster may thus operationally 
``displace'' nearby objects, if any, making them appear farther 
than they truly are. Figure \ref{histos.ps}(e) 
suggests that such may indeed be the case, as the number of objects
plotted between $10^{7.5}$ and $10^8$ $M_\odot$ is overabundant for
any reasonable shape of the HI mass function, in comparison with the
higher mass bins. Redshift--independent
distance estimates of those objects are necessary, if a reliable
determination of the HI mass function in Virgo and in the field is to
be obtained. This issue will be addressed in a forthcoming paper by
Kent \etal ~(2007, in preparation).

Figure \ref{himdist.ps}(a) displays a plot of HI mass versus distance
for all objects in Table 1. Vertical dotted lines outline the distance
interval within which our survey is unable to detect any cosmic sources,
due to the intrusion of rfi. Galaxies in the Virgo cluster appear as the
vertical feature 
near $D=17$ Mpc. Figure \ref{himdist.ps}(b) shows that the S/N limit
of the survey is relatively independent of velocity width. Sources
with velocity widths larger than a few hundred \kms ~are apparently 
identified reliably by our signal extraction algorithm. Very few objects
are found with velocity width smaller than 25 \kms. Several candidate
detections are obtained with narrow widths; they lie generally near the
detection limit and have no identifiable optical counterparts. Most of
them are believed to be due to rfi. Determination of their nature will
require follow--up corroborating observations.

Figure \ref{himdist.ps}(c) shows the dependence of the survey's limiting
flux integral on velocity width. The dotted line indicates a detection
threshold of $S/N=6.5$, assuming that a spectral smoothing width
of $W50/20$ can be adopted for features as wide as 400 \kms, and 
a constant value of $400/20$  for features wider than 400 \kms
(see description of Col. 8 of Table 1).
The fact that the lower envelope of the data points appears 
consistent with a slope of 1/2, rather than steepening to a slope
of 1, indicates that our adoption of a smoothing width is 
essentially correct.

\section {Positional Accuracy of HI Sources \label{pos}}

The main purpose of this work is to make available to the general
community a data set of HI detections, with analysis of the
overall properties of the sample kept at a minimum in the interest of
speedy delivery. We address the reader to Paper I and Paper II, in which a
preliminary analysis was discussed. However, because of the dependence
of the performance properties of the Arecibo telescope on the direction
in which it points, some characteristics of the ALFALFA samples will
be sky zone dependent. One of such characteristics is the positional
accuracy of the HI detections, which carry the imprint of the
telescope pointing errors. Thus we devote this section to a brief
analysis of that matter. The corrections we will find are specific
to the data set presented here.

Positional accuracy of HI sources is of paramount importance per se and
in making identification with sources from other catalogs. The main
limiting factor for ALFALFA sources is of course the resolution of the
ALFA beam. As discussed in Paper I, the ALFA beams are slightly elliptical,
with half-power-full-widths of 3.3\arcmin ~and 3.8\arcmin. The major axis of the
beam is always directed along the position of the telescope's azimuth
arm; the vast majority of the ALFALFA observations and all those
presented in this paper are made with the azimuth arm in the North-South 
direction, thus the beam's major axis is in the Declination direction.

ALFALFA samples the sky every second in R.A. and every arcminute in Dec.
However, sources are extracted --- and their positions measured --- after
the data are converted to a spatial grid sampled at  $1$\arcmin ~$\times$ $1$\arcmin. 
A Gaussian weight 
function is applied as part of the regridding process, which reduces the spatial 
resolution of the data to $3.8$\arcmin ~$\times 4.3$\arcmin. An automatic source extraction
algorithm identifies a source candidate, which is successively measured
and ``extracted'' interactively by an observer. The spectral extent of
the feature is gauged and a 2--D map of the emission is integrated over
that full spectral extent. Ellipses are fitted to the image at a set of 
fixed isophotal levels as well as at the half--power and quarter--power
of the peak level, as measured in flux density units per beam area. The
position of the source is assumed, in all cases, to be the center of the 
half-power ellipse. This has proved to be the best choice for the vast
majority of sources, as they are generally unresolved by the ALFA beam.
Caution is necessary when sources are extended and do not display a
clear center of symmetry. A more detailed description of the positional 
parameter estimates will be presented elsewhere, together with the overall 
description of the ALFALFA data processing pipeline (Giovanelli \etal ~2007,
in preparation).

The second most important parameter regulating the quality of the ALFALFA 
positions is the S/N of the HI emission. High S/N sources
allow more accurate centroiding than low S/N ones, as discussed below.

The third important influence on ALFALFA positions is the quality of the
Arecibo telescope pointing. The telescope can set with a repeatable
accuracy of a few arcsec; however, the pointing algorithms which use fits
to the telescope configurational parameters yield systematic pointing
errors which may, at L band, add up to 15\arcsec ~or more. The highest pointing 
errors occur at the lowest zenith angles. The chosen observing
mode for ALFALFA, which freezes the telescope configuration at fixed
azimuth, largely allows recovery and correction of the systematic pointing
errors.

Positional accuracy is discussed here by using the positional differences
between the HI emission ellipse centers and the centers of the galaxies
which are identified as optical counterparts. Those differences will include
systematic telescope pointing offsets, occasional mismatch in the extent and
center of optical and HI source, statistical errors in the HI centroiding and
misidentification of optical counterparts. Errors in the centroiding of the
optical sources are negligible in comparison. A similar exercise to the one
described below has been
carried out by comparing ALFALFA positions of continuum radio sources with
interferometic positions. Those results are in agreement with those discussed
here and will be presented elsewhere (Kent et al. 2007, in preparation).

%FIGURE X
%\begin{figure}[ht!]
%\epsscale{0.7}
%\plotone{figs/dra_vs_ddec.ps}
%\caption{.}
%\label{dravsddec}
%\end{figure}
% This is the figure labeled dra_vs_ddec.ps

%FIGURE 6
\begin{figure}[ht!]
%\figurenum{1}
\plotone{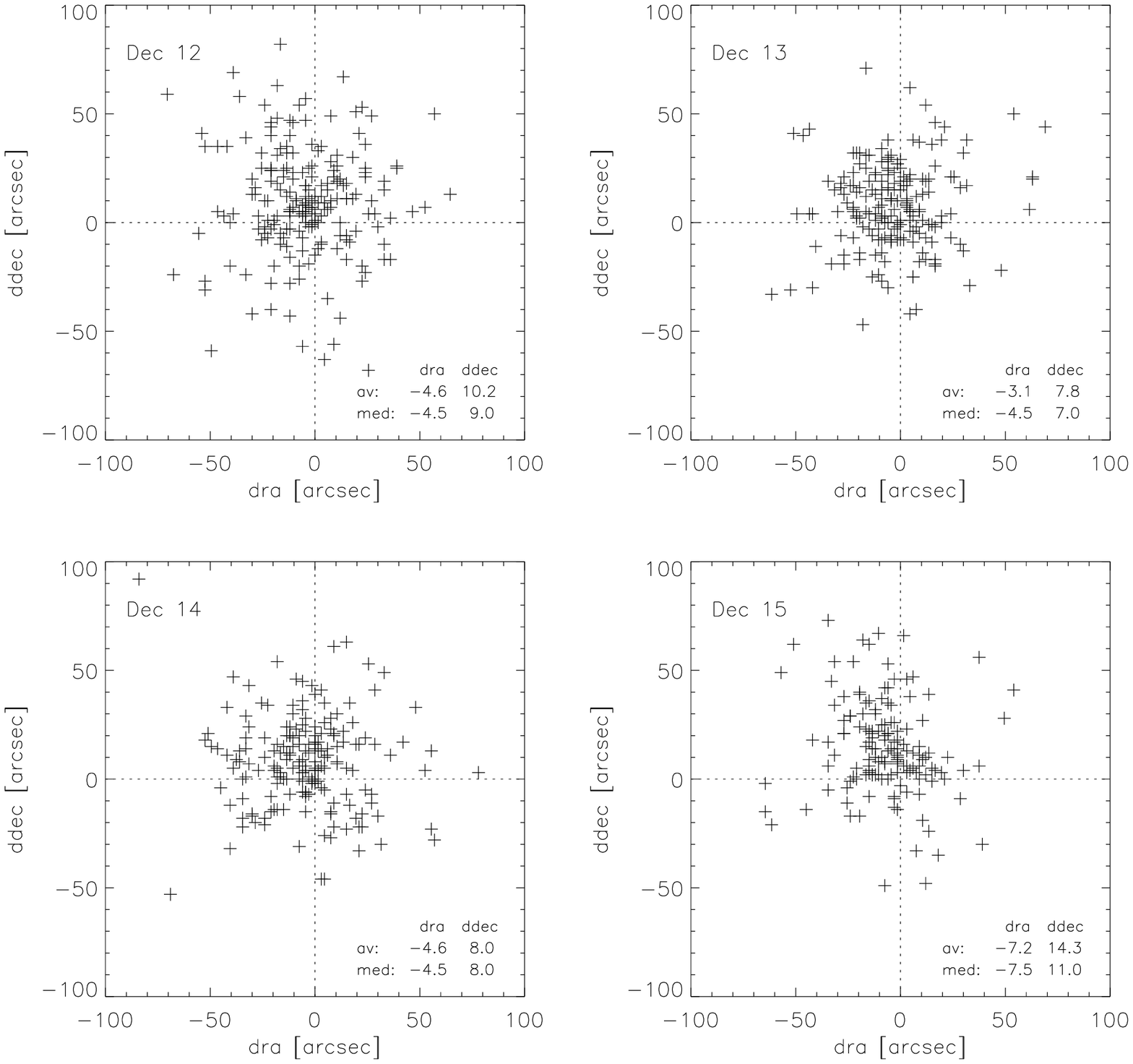}
\caption{Differences between the positions of the HI sources, {\it before
applying a correction for telescope pointing errors}, and the
optical counterpart positions as listed in Col. 4 of Table 1. Sources
are separated by Declination bins of $1^\circ$: the label ``Dec 12'' identifies 
sources with Declination between +12$^\circ$ and +13$^\circ$, etc. Average 
and median offsets, expressed in arcsec, are inset within each panel.}
\label{dravsddec4}
\end{figure}

Figure \ref{dravsddec4} shows the positional differences (HI minus optical), 
respectively in R.A. ($\Delta ra$) and in Dec. ($\Delta dec$), expressed 
in arcsec. The data are split into four panels differing in Declination
range, within bins of  $1^\circ$. The frame labeled ``Dec 12''
refers to sources between Dec = $+12^\circ$ and Dec = $+13^\circ$ and so on.
As the observations were made with the telescope's azimuth arm oriented
N--S, a degree of Declination converts exactly in 1 degree of zenith
angle; the latitude of the Arecibo telescope is 
$+18^\circ$ 20\arcmin ~37\arcsec ~N.  The plot includes all the
sources presented in Table 1, for which an optical identification was made.
The HI positions used are those obtained {\it before} any correction for telescope pointing errors
was applied. These sources were all observed at the same azimuth of $0^\circ$ 
and zenith  angles between $2.3^\circ$ and $6.4^\circ$.
A systematic offset in the center of the distribution towards positive 
$\Delta ra$ and negative $\Delta dec$ is apparent in all panels, with
that offset becoming progressively larger with increasing Declination,
i.e. decreasing zenith angle. This offset mimics the telescope pointing errors
(see ``pointing errors'' at the website 
{\it http://www.naic.edu/$\sim$phil/sysperf/sysperf.html\#alfa})
which increase as the telescope points closer to the site's zenith. The
amplitude of the pointing errors is indicated by the insets in each plot,
expressed respectively by the mean and the median.

%FIGURE 7
\begin{figure}[ht!]
%\figurenum{1}
\plotone{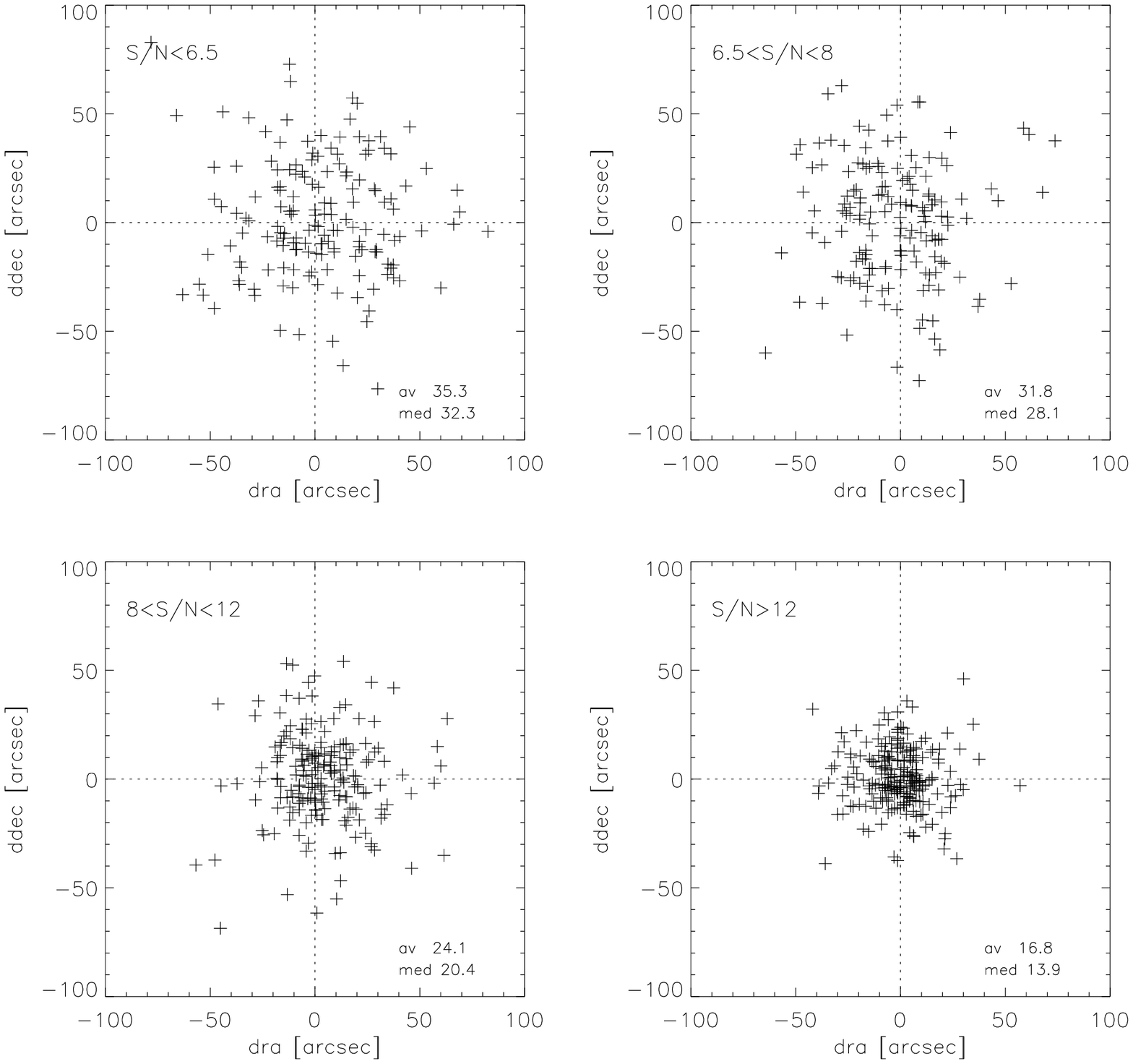}
\caption{Differences between the positions of the HI sources, as indicated by
the values in Col. 3 of Table 1, which are corrected for systematic telescope
pointing errors, and the
optical counterpart positions as listed in Col. 4 of Table 1. Sources
are separated by S/N as indicated in each panel. Average and
median offsets, expressed in arcsec, are inset within each panel.
}
\label{dravsddec4stn}
\end{figure}

Figure \ref{dravsddec4stn} shows the distribution of positional offset between 
HI and optical positions after the systematic pointing errors described above 
have been removed from the HI positions.
The separation of the data into four panels in this case is by S/N.
The systematic pointing error that was removed is a simple function of Dec., i.e.
zenith angle, obtained for each source by interpolating between the offsets shown 
in Figure \ref{dravsddec4}.
The HI source positions corrected for this systematic error are those listed in 
column 2 of Table 1. The dispersion
of the positional differences about the center is, as expected, dependent on S/N.
For the higher S/N objects ($S/N>12$), ALFALFA positions are on average accurate
to about 15\arcsec. ALFALFA positions are significantly better than those obtained
with the precursor run reported by Paper II (median difference between HI and optical
position of 34\arcsec). This is due to two reasons: (i) the ALFALFA data include two
full sweeps of each region of the sky with ALFA, hence yielding twice as dense a
spatial sampling in Dec. as most of the data in the precursor run; (b) the technique
used to extract positional information, which fits ellipses to isophotal contours,
is more accurate than that used for the precursor run data.

%FIGURE X
%\begin{figure}[ht!]
%\epsscale{0.7}
%\plotone{figs/dra_vs_ddec_corr.ps}
%\caption{.}
%\label{dravsddec4corr}
%\end{figure}

\section {New Discoveries, Future Work and Conclusions}\label{disc}

The principal aim of ALFALFA is to obtain an accurate census of HI-bearing objects
in the local universe. The catalog of candidate HI detections presented in Table 1
yields a first picture of the ``ALFALFA sky''. As discussed in Section \ref{stats} and
shown in Figure 4, ALFALFA detections span five orders of magnitude in HI mass,
and include both massive spirals out to $z \sim$ 0.06 and dwarf galaxies
within a few Mpc of the Milky Way. The region sampled by the present paper is
strongly affected by local large scale structure (the Virgo Cluster), as evident 
in the redshift distribution presented in Figures \ref{histos.ps} ~and \ref{cone.ps}. 
Detailed studies regarding environmental influences on HI content, the HIMF and other 
characteristics will be incrementally enabled as further installments of ALFALFA 
data become available.

The principal conclusions which can be gleaned from this first installment of
ALFALFA sources in a region covering 132 \sqd, which represents
only 1.9\% of the final survey, are:

\begin{itemize}
\item
The ALFALFA survey is delivering high quality 1.4 GHz spectral line data as anticipated.
The adopted ``minimum intrusion'' observing technique (Paper I) is highly successful at reducing
spectral baseline instability, beam and sidelobe variations and gain instabilities.
Through a combination of deliberate calibration technique and empirical
checks, the positional and photometric accuracy of extracted sources is
meeting or even exceeding anticipated survey design specifications.

\item 
Although this region of the sky has been previously targeted heavily
by surveys of optically selected galaxies (Gavazzi \etal ~2003) and was covered by 
both the HIPASS (Wong \etal ~2006) and HIJASS (Davies \etal ~2004) Virgo cluster
surveys, 69\% of the HI detections
presented here are new. The improvement over HIPASS in this region is
a factor of 18 in the number density of HI detections.
The galaxies detected by ALFALFA include a 
population of gas-rich, star-forming low surface brightness galaxies
which are not included in previous optical magnitude-limited surveys.

\item 
The median redshift of ALFALFA in this region of the sky is $\sim$ 7000 \kms,
This depth may be compared to that of HIPASS: $\sim$ 2800 \kms (Meyer \etal ~2004).
ALFALFA samples volumes well beyond the Local Supercluster, and 
the distribution of HI detections follows the large scale structure evident
in the region. Eventually, we will use the ALFALFA dataset to measure
cosmological parameters such as the clustering properties of the HI population 
and its bias parameter, and to explore further the ``void problem'' (Peebles
2001).

\item
27 of the HI sources are identified with galaxies of early morphology, types
E, dE/Sph or S0. Half are in the vicinity of the Virgo cluster and could be members.
A study of the characteristics of these objects and the morphology and 
kinematics of the HI is under way by Koopmann \etal ~(2007, in preparation).

\item
Among the objects tabulated as HI detections here, several appear
to be extended complexes of HI clouds. Referring to them by their AGC number
(cat. nr.),
226054 (1-235), 226055 (1-238), 226056 (1-239) and 224316 (1-241)
are different clumps of the ``dark galaxy'' Virgo HI21 reported by Davies 
\etal ~(2004).
The ALFALFA observations show that this feature is clearly connected to
the nearby one-armed spiral NGC~4254; 
we discuss these observations in more detail elsewhere (Haynes \etal ~2007,
in preparation).

\item
For \nooc ~HI sources, we have been unable to identify unambiguously their
optical counterparts. More than half of these are found at redshifts less than
$+200$ \kms ~and are likely to be perigalactic HVCs. Particularly notable 
among them are the compact HVCs with positive
velocities, a peculiar population in the vicinity of the North Galactic pole.
A number of the objects reported by de Heij, Braun \& Burton (2002) with fluxes close
to the detection limit of that survey are not confirmed by the more sensitive 
ALFALFA observations (Kent \etal ~2007, in preparation). The objects with positive 
velocity which are confirmed, and a few previously undetected with similar velocities, 
lie in the same sky region as several distant perigalactic structures exhibiting 
positive velocities: the Bootes dwarf spheroidal galaxy (Belokurov \etal ~2006a; 
Mu\~noz \etal ~2006),  the leading arm of the Sagittarius stream and 
others found in the the ``Field of Streams'' (Belokurov \etal ~2006b), and farther away but still
at $\sim$200 \kms, the Local Group galaxy GR~8. We are in the process of investigating
the interpretation of these features in more detail.

\item
For some of the remaining objects without clear optical identifications, 
possible optical counterparts exist, but the positional accuracy of the ALFALFA
HI data is insufficient to yield an unambiguous identification. Interestingly, 
17 of the candidate HI detections with no discernible optical 
counterpart are unlikely to be HVCs. In the case of 215217 (1-62), 
226061 (1-257), 226080 (1-316)
and 223449 (1-271), galaxies are seen within few arcminutes of the HI positions, 
which have redshifts comparable with those of the HI features. AGC 226080
has been shown by Oosterloo \& van Gorkom (2005) to be a plume extending from
NGC~4388. AGC 223449 may have
very low HI mass, although its distance is highly uncertain. It remains to be
verified whether a connection between the optical and HI objects exists.
In the case of 225998 (1-173), another object of highly uncertain distance,
a HI--rich galaxy (U7235=NGC4189, 1-175) is found at comparable redshift,
but half a degree away.
AGC 226118 (1-354) could be an interesting feature in the Virgo cluster, but its
vicinity to M87 makes the detection doubtful in spite of a better than fair
S/N. AGC 215230 (1-44), 226043 (1-185), 226117 (1-352), 226119 (1-360)
nd 226120 (1-366) have velocities in excess of 4000 \kms;
most in that group are marginal candidate detections. All others sources
have identified optical counterparts, thanks to the good positional accuracy
of ALFALFA source candidates. A preliminary analysis of Virgo cluster HI sources 
with no optical counterparts is given in Kent \etal (2007, in preparation).

\end{itemize}

In summary, we present the first catalog release of the ALFALFA survey,
corresponding to 1.9\% of the projected sky coverage of the completed survey.
As anticipated, ALFALFA delivers
a dramatic improvement in HI detection sensitivity over previous HI blind
surveys through its combination of wide areal coverage, smaller
beam area, higher spectral resolution and the sheer sensitivity superiority
offered by Arecibo's huge collecting area. 
The vast majority of HI sources listed in the present catalog have identified 
optical counterparts. Many are vigorously star-forming yet
optically faint, late-type galaxies. Among those HI sources which cannot be 
unambiguously identified with an optical galaxy, we 
find a population of HVCs, including ones at significant positive velocity,
and a few legitimately extragalactic objects  whose optical counterparts are 
not yet identified.

In addition, the survey identifies nearly as many candidate detections of lower 
S/N, including many with narrow spectral lines and no optical counterparts.
Because candidate source reliability plummets below S/N $<$ 5, 
these HI candidates are not reported here but are targeted for corroborating 
follow--up observations with the Arecibo L-band wide system employing an
efficient strategy designed for that purpose, as discussed in Paper I.

A catalog of comparable size, covering a strip of the same R.A. extent 
and including the southern part of the Virgo cluster, is in 
preparation (Kent \etal ~2007, in preparation); eventually, ALFALFA
will cover the entire region included in the Virgo Cluster Catalog
(Binggeli, Sandage \& Tammann 1985). Catalog data releases for other parts
of the sky are also in preparation by several groups within our collaboration,
as are programs involving multiwavelength follow--up studies of
selected targets presented here.

The catalog presented here and its associated data products will be incorporated
into the more extensive digital HI dataset at
{\it http://arecibo.tc.cornell.edu/hiarchive} as part of the ALFALFA
Arecibo legacy.

\vskip 0.3in
 
RG, MPH, NB, TB and RK acknowledge the partial support of NAIC as Visiting 
Scientists during the period of this work. This work has been supported by NSF 
grants AST--0307661, AST--0435697, AST--0347929, AST--0407011, AST--0302049;
and by a Brinson Foundation grant. We thank the Director of
NAIC, Robert Brown, for stimulating the development of major ALFA surveys, 
H\' ector Hern\' andez for his attention to the telescope scheduling and the Director, 
telescope operators and support staff of the Arecibo Observatory for their proactive 
approach. We also thank Tom Shannon for his advice and assistance with hardware, 
system and network issues at Cornell.

\vfill

\end{document}